\documentclass[aps,prd,twocolumn,showpacs,amsmath,amssymb]{revtex4}
\usepackage{graphicx,color,dcolumn,booktabs,bm}
\usepackage{longtable,lscape,comment,braket}
\usepackage{txfonts}
\usepackage{overpic}
\usepackage{amssymb}
\usepackage{indentfirst}
\usepackage{feynmf}   
\usepackage{slashed}  
\usepackage{cases}
\usepackage{epstopdf}
\usepackage{psfrag}
\usepackage{subfigure}
\usepackage{threeparttable}
\pdfoutput=1
\usepackage{color}
\usepackage{multirow}
\usepackage{url}

\graphicspath{{Figures/}} %
\usepackage[colorlinks, citecolor=blue,anchorcolor=red,menucolor=red, linkcolor=red,filecolor=red,runcolor=red,urlcolor=blue,frenchlinks=red]{hyperref}

\usepackage{ulem}
\def\be{\begin{equation}}
\def\ee{\end{equation}}

\begin{document}

\title{Charged charmoniumlike structures in the $e^+ e^- \to \psi(3686) \pi^+ \pi^-$ process based on the ISPE mechanism}

\author{Qi Huang$^{1,2}$}\email{huangq16@lzu.edu.cn}
\author{Dian-Yong Chen$^3$}\email{chendy@seu.edu.cn}
\author{Xiang Liu$^{1,2}$\footnote{Corresponding author}}\email{xiangliu@lzu.edu.cn}
\author{Takayuki Matsuki$^{4,5}$}\email{matsuki@tokyo-kasei.ac.jp}

\affiliation{
$^1$School of Physical Science and Technology, Lanzhou University, Lanzhou 730000, China\\
$^2$Research Center for Hadron and CSR Physics, Lanzhou University and Institute of Modern Physics of CAS, Lanzhou 730000, China\\
$^3$School of Physics, Southeast University, Nanjing 210094, China\\
$^4$Tokyo Kasei University, 1-18-1 Kaga, Itabashi, Tokyo 173-8602, Japan\\
$^5$Theoretical Research Division, Nishina Center, RIKEN, Wako, Saitama 351-0198, Japan}

\begin{abstract}

In 2017, the BESIII Collaboration announced the observation of a charged charmonium-like structure in the $\psi(3686)\pi^\pm$ invariant mass spectrum of the $e^+ e^- \to \psi(3686) \pi^+ \pi^-$ process at different energy points, which enables us to perform a precise study of this process based on the initial single pion emission (ISPE) mechanism. In this work, we perform a combined fit to the experimental data of the cross section of $e^+ e^- \to \psi(3686) \pi^+ \pi^-$, and the corresponding $\psi(3686)\pi^\pm$ and dipion invariant mass spectra.
Our result shows that the observed charged charmonium-like structure in $e^+ e^- \to \psi(3686) \pi^+ \pi^-$ can be well reproduced based on the ISPE mechanism, and that the corresponding dipion invariant mass spectrum and cross section can be depicted with the same parameters. In fact, it provides strong evidence that the ISPE mechanism can be an underlying mechanism resulting in such novel a phenomenon.

\end{abstract}


\maketitle

\section{Introduction}
\label{Sec:Intro}

Since 2003, study on the charmoniumlike $XYZ$ states has become a hot spot of hadron physics and particle physics. Since it has
the close relation to non-perturbative behavior of quantum chromodynamics (QCD), the relevant investigation of the charmoniumlike $XYZ$ states
is helpful to deepen our understanding of strong interaction. In the past sixteen years, there have been extensive discussions about this issue (see reviews \cite{Liu:2013waa,Chen:2016qju,Liu:2019zoy,Guo:2017jvc} for more details).

To solve the puzzling phenomena existing in the hidden-bottom dipion decays of $\Upsilon(10860)$, the initial single pion emission (ISPE) mechanism was proposed in Refs. \cite{Chen:2011pv,Liu:2011ky,Chen:2012yr}, where the charged $Z_b(10610)$ and $Z_b(10650)$ structures can be qualitatively reproduced. In 2012, considering the similarity between the bottomonium and charmonium families, the ISPE mechanism was applied to study the hidden-charm dipion decays of higher charmonia and charmoniumlike state $Y(4260)$, where the predicted charged charmoniumlike structures near $D\bar{D}^*$ or $D^*\bar{D}^*$ threshold may exist in the corresponding $J/\psi\pi^+$, $\psi(3686)\pi^+$, and $h_{c}\pi^+$ invariant mass spectra \cite{Chen:2011xk}. These predictions given by the ISPE mechanism have provided valuable information to search for such charged charmoniumlike structures.

In 2013, the BESIII and Belle collaborations indeed discovered a charged charmonium-like $Z_c(3900)$ in the process $e^+ e^- \to J/\psi \pi^+ \pi^-$ around $\sqrt{s}=4.26\ \mathrm{GeV}$ \cite{Ablikim:2013mio,Liu:2013dau}, which was quickly and further confirmed by CLEO-c in the same process but at $\sqrt{s}=4.17$ GeV \cite{Xiao:2013iha}.
Later, another charged charmoniumlike structure named $Z_c(4020)$ was observed in the $h_c \pi^+$ invariant mass spectrum of the $e^+ e^- \to \pi^+ \pi^- h_c$ process \cite{Ablikim:2013wzq}. Additionally, at the center-of-mass energy of 4.26 GeV, the BESIII Collaboration reported two charged charmoniumlike structures, $Z_c(3885)$ in the $(D^\ast \bar{D})^{\pm}$ invariant mass spectrum of $e^+e^- \to \pi^\mp (D^\ast \bar{D})^\pm$ process \cite{Ablikim:2013xfr} and $Z_c(4025)$ in the $(D^\ast \bar{D}^\ast)^\pm$ invariant mass spectrum of the $e^+ e^- \to \pi^\mp (D^\ast \bar{D}^\ast)^\pm$ process \cite{Ablikim:2013emm}.

\begin{figure}[htbp]
\begin{center}
\scalebox{0.32}{\includegraphics{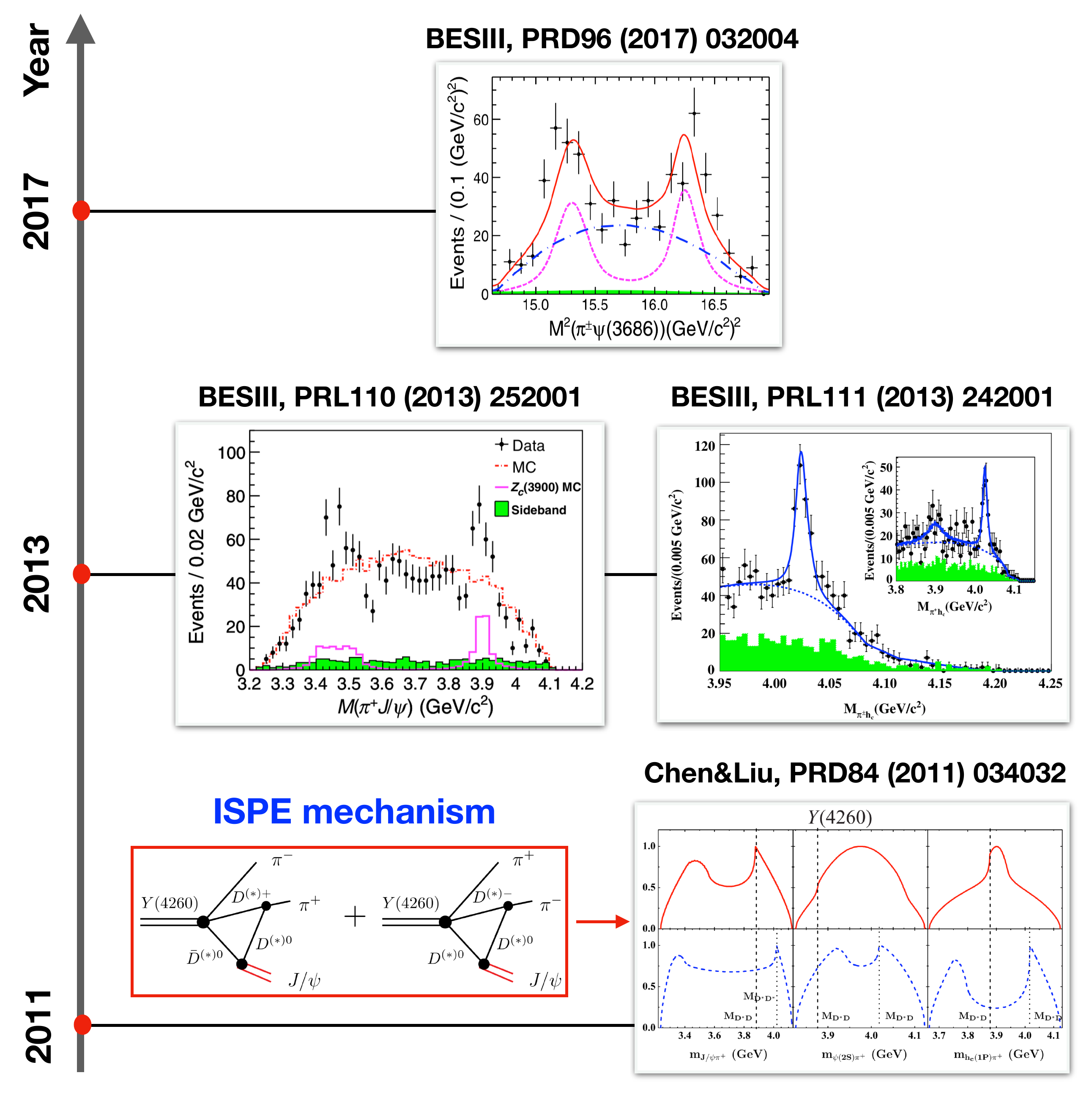}}
\caption{The predicted charged charmoniumlike structures appearing in $J/\psi\pi^\pm$, $h_c\pi^\pm$ and $\psi(2S)
\pi^\pm$ invariant mass spectrum and the corresponding experimental discoveries \cite{Ablikim:2013mio,Ablikim:2013wzq,Ablikim:2017oaf}. \label{ispe}}
\end{center}
\end{figure}

The story of finding charged charmoniumlike structures is continued. In 2017, the BESIII experiment made a progress on observing charged charmoniumlike structure in the $ \psi(3686)\pi^+$ invariant mass spectrum
by analyzing the data of cross section of the
$e^+e^-\to \psi(3686)\pi^+\pi^-$ process at different energy points $\sqrt{s}=4.226,\,4.258,\,4.358,\,4.387,\,4.416,$ and $4.600$ GeV \cite{Ablikim:2017oaf}.
In fact, this observation deepens our confidence again. As shown in Fig. \ref{ispe}, the predicted charged charmoniumlike structures in the $J/\psi\pi^+$, $\psi(3686)\pi^+$, and $h_{c}\pi^+$ invariant mass spectrum have been reported in experiments until now.

The BESIII's experimental measurement of the $e^+e^-\to \psi(3686)\pi^+\pi^-$ process forces us to carry out a further study of this process based on the ISPE mechanism. Before the present work, in Ref. \cite{Chen:2013coa}, the similar idea was once adopted to study the $e^+e^-\to J/\psi\pi^+\pi^-$ process combined with the BESIII data of the $J/\psi\pi^\pm$ invariant mass spectrum and the corresponding dipion invariant mass spectrum
of $e^+e^-\to J/\psi\pi^+\pi^-$ at one energy point $\sqrt{s}=4.26$ GeV, which shows that the $Z_c(3900)$ structure can be reproduced well by the ISPE mechanism. Different from Ref. \cite{Chen:2013coa}, the present work will perform a combined fit to the released data of the $\psi(3686)\pi^\pm$ and $\pi^+\pi^-$ invariant mass spectrum from the $e^+e^-\to \psi(3686)\pi^+\pi^-$ process at different energy points. Comparing with the former work \cite{Chen:2013coa}, the present work is obviously a tough task and is full of challenge. In this work, we will reproduce the structures observed in the $e^+e^-\to \psi(3686)\pi^+\pi^-$ process step by step based on the ISPE mechanism. It also reveals the relation between the
observed charged charmoniumlike structures and
the ISPE mechanism, which will be helpful to deepen our understanding on such a novel phenomenon.

This work is organized as follows. After introduction, we perform an analysis of the mechanisms working in the $e^+ e^- \to \psi(3686) \pi^+ \pi^-$, and then fit the present model with the experimental data. In Sec. \ref{Sec:Num}, we present our fit results to the cross sections for $e^+ e^- \to \psi(3686) \pi^+ \pi^-$ and the mass spectra of $\psi(3686) \pi^\pm$ and dipion invariant mass spectra. Sec. \ref{Sec:Sum} is devoted to conclusions and discussion.

\section{Mechanisms occurring in $e^+ e^- \to \psi(3686) \pi^+ \pi^-$}
\label{Sec:Mec}

\begin{figure}[t]
\centering
\begin{tabular}{cc}
  \scalebox{0.285}{\includegraphics{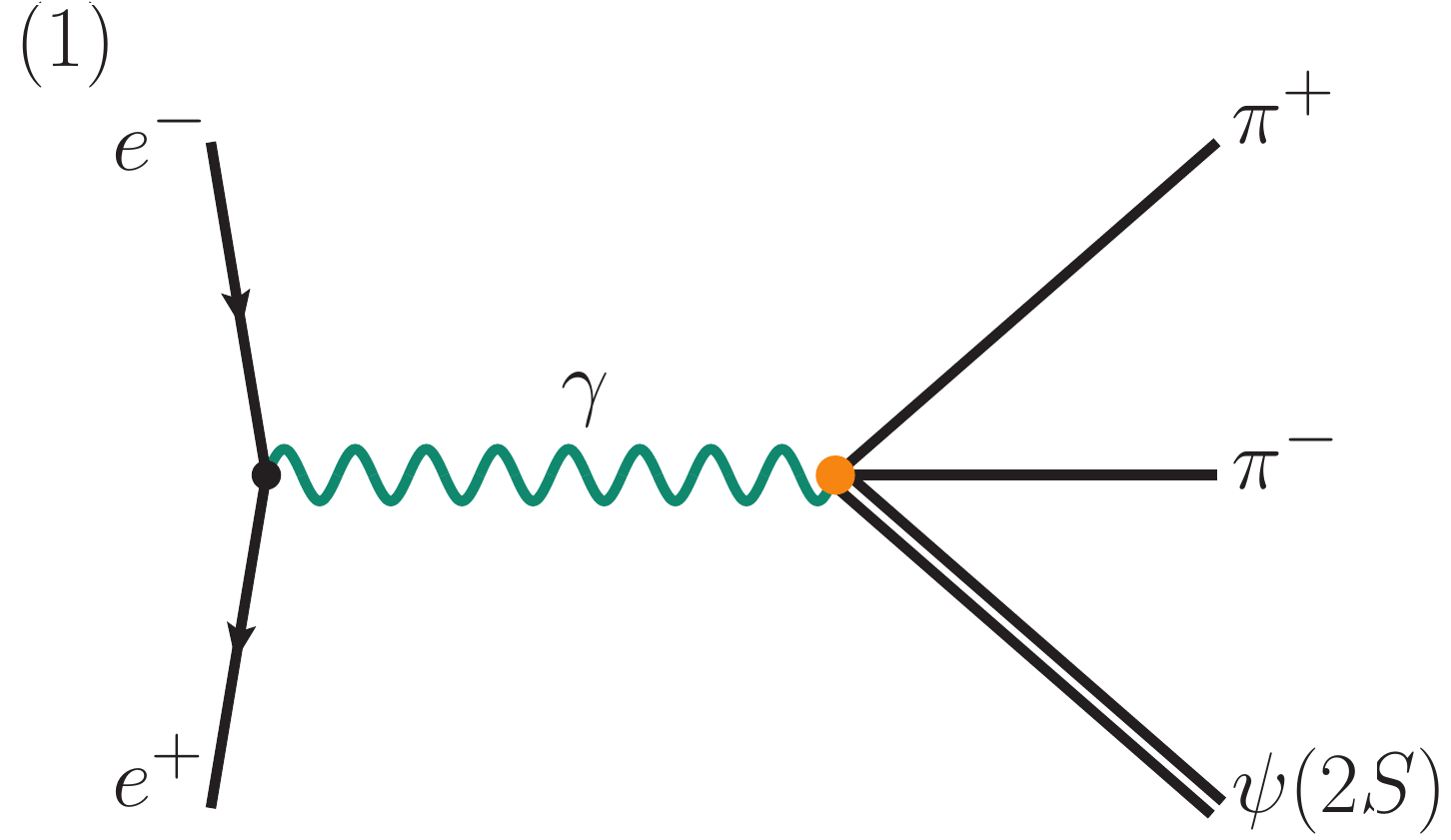}}  &
  \scalebox{0.285}{\includegraphics{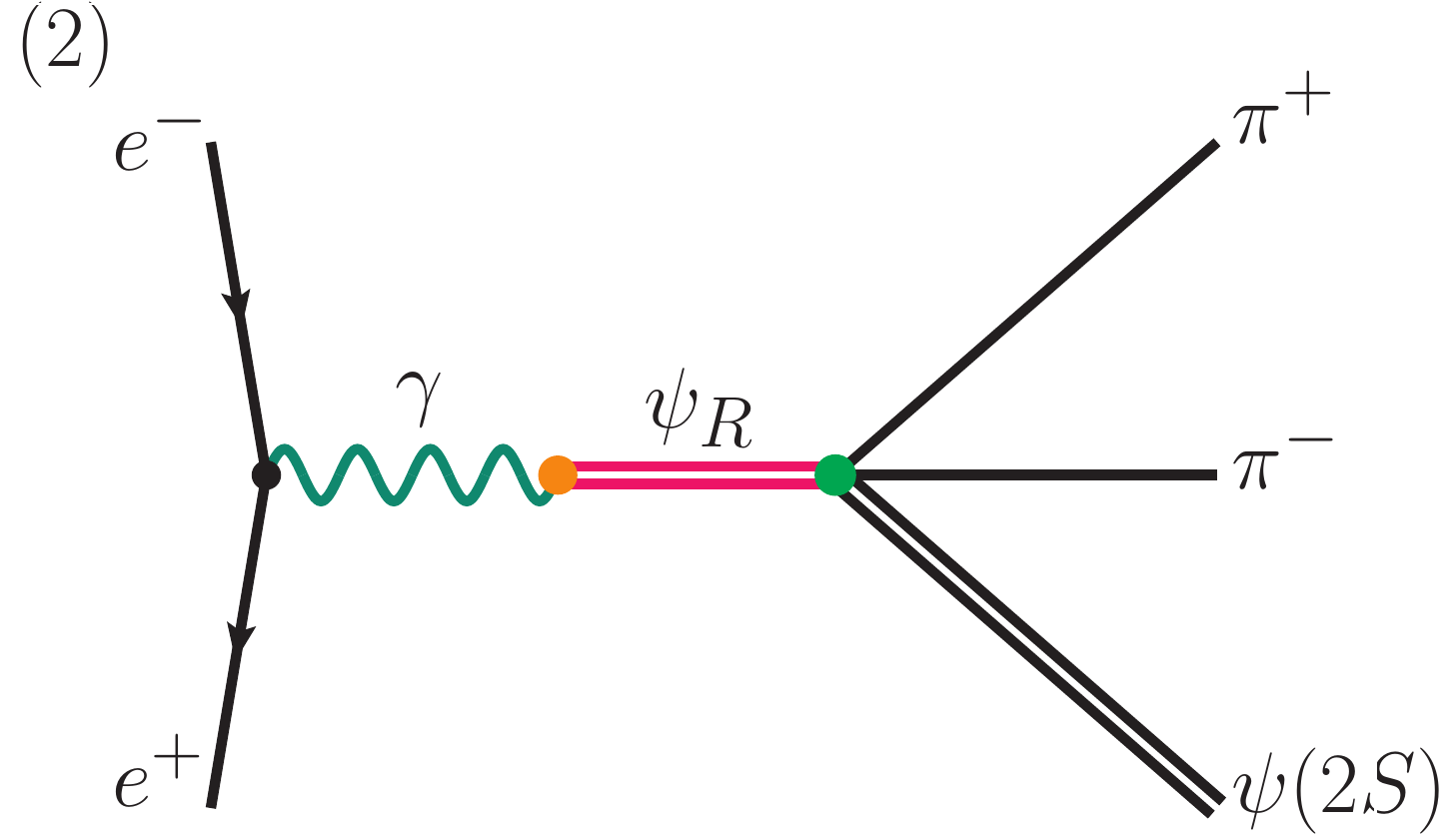}}\\
  \scalebox{0.285}{\includegraphics{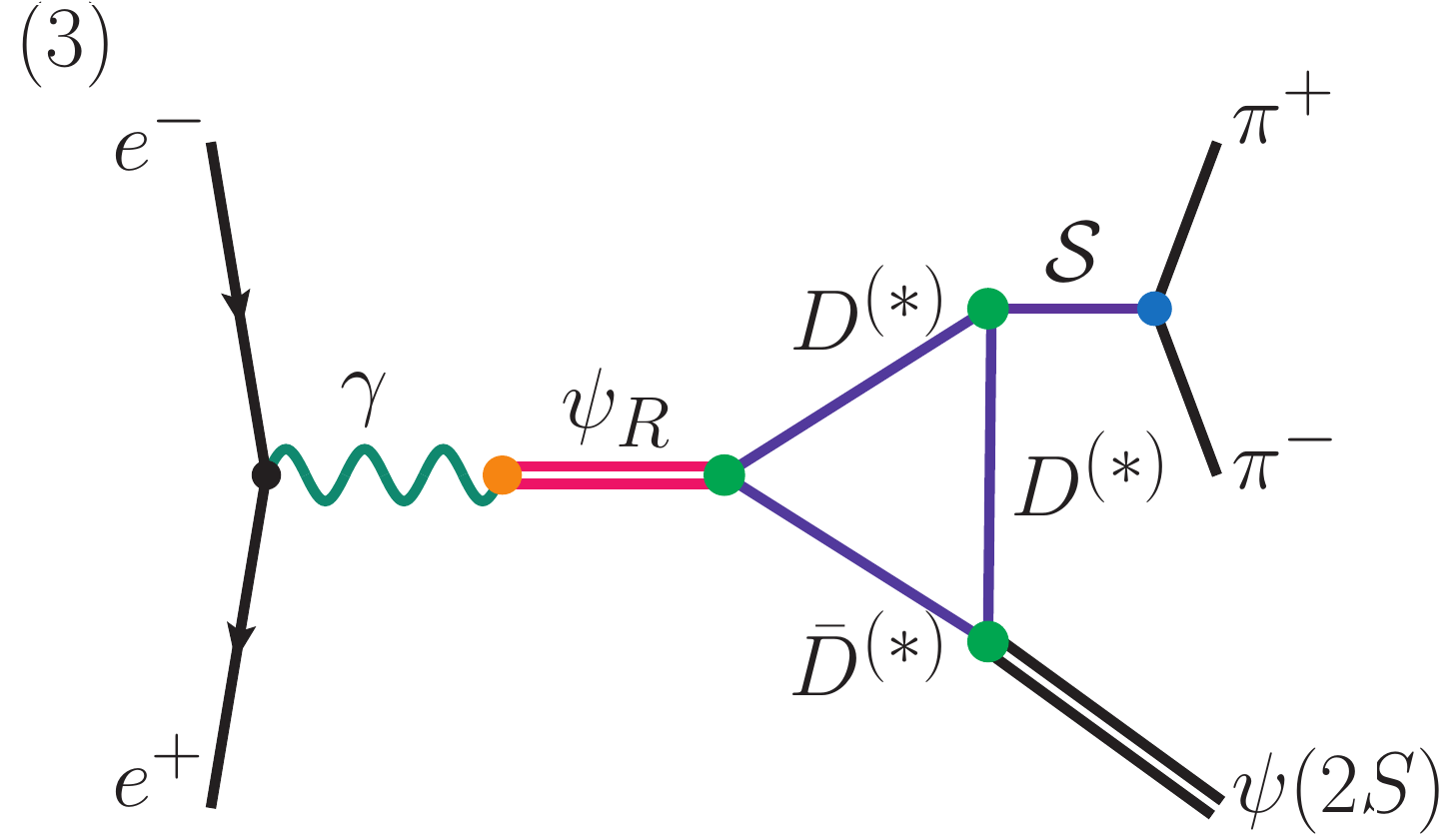}} &
   \scalebox{0.285}{\includegraphics{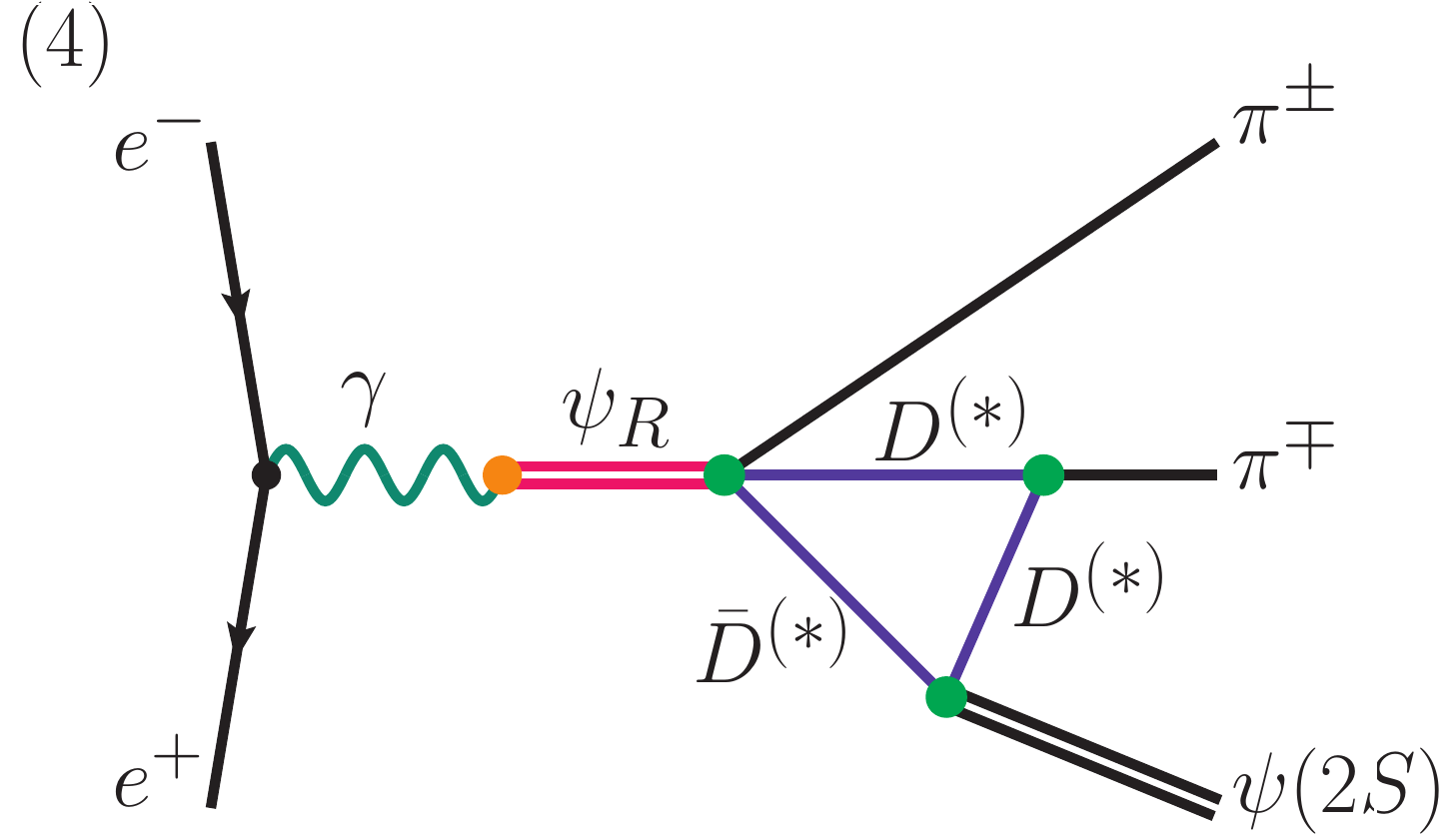}}
\end{tabular}
 \caption{(Color online). Feynman diagrams depicting the decay mechanisms which give contributions to the process $e^+ e^- \to \psi(3686) \pi^+ \pi^-$.}
  \label{Fig:Mech}
\end{figure}

In Ref. \cite{Chen:2013coa}, we once reproduced the $Z_c(3900)$ structure
in the $e^+ e^- \to J/\psi \pi^+ \pi^-$ process based on the ISPE mechanism. In the present work, we will apply the same idea to
discuss $e^+ e^- \to \psi(3686) \pi^+ \pi^-$. There exist serval different decay mechanisms occurring in the process $e^+ e^- \to \psi(3686) \pi^+ \pi^-$ shown in Fig. \ref{Fig:Mech}. Fig. \ref{Fig:Mech} (1) represents the nonresonance contribution, where the final states $\psi(3686)\pi^+ \pi^-$ directly couple to the photon produced by the electron-positron annihilation. Fig. \ref{Fig:Mech} (2)-(4) correspond to the processes with an intermediate charmonium resonance, where the virtual photon first couples to a vector charmonium $\psi_{R}$, which decays into $\psi(3686)\pi^+\pi^-$. There are three different kinds of decay mechanisms in the higher charmonium dipion decays: i) As shown in Fig. \ref{Fig:Mech} (2), the first is the one in which the dipion is produced by the gluon hadronization with gluons emitted by a charm quark. ii) In the second decay mechanism, the dipion is produced through a scalar meson decay, while the scalar meson and $\psi(3686)$ are connected to the initial $\psi_{R}$ by a charmed meson loop. This is presented in Fig. \ref{Fig:Mech} (3). iii) The last decay mechanism is the ISPE mechanism, which is presented in Fig. \ref{Fig:Mech} (4). We will list resonaces $R$ inserted into Fig. \ref{Fig:Mech} later in Table \ref{tab:psi-parameter}.

In the following, we will present the amplitudes corresponding to different mechanisms. For Fig. \ref{Fig:Mech} (1), its amplitude is expressed as
\begin{eqnarray}
\mathcal{M}^{(1)}_{\rm NoR}=\bar{v}(k_2) \gamma^\mu u(k_1) g_{\rm NoR} e \frac{-g_{\mu\nu}}{s} \epsilon^{\ast \nu}_{\psi^\prime} \mathcal{F}_{\rm NoR}(s),\label{eqs:background}
\end{eqnarray}
where $g_{\rm NoR}$ is a coupling constant, $e=\sqrt{4\pi\alpha}$ denotes an electric charge with $\alpha=1/137$ being the fine structure constant, and $\mathcal{F}_{\rm NoR}(s)$ is a form factor. In the present work, we adopt the following form factor \cite{Chen:2013coa}
\begin{eqnarray}
\mathcal{F}_{\rm NoR}(s)=  e^{-a_{\rm NoR}(\sqrt{s}-\Sigma m_f)^2},\label{eqs:ff}
\end{eqnarray}
where $a_{\rm NoR}$ is a phenomenological parameter, which should be fixed when fitting the data. $\Sigma m_f=m_{\psi^\prime}+2m_{\pi^\pm}$  is the sum of the masses of the involved particles appearing in the final states.

For Figs. \ref{Fig:Mech} (2)-(4), a general amplitude, with a help of the vector meson dominance (VMD) ansatz \cite{Lin:2013mka,Xu:2015qqa}, can be written as
\begin{eqnarray}
\mathcal{M}^{(i)}_{R}&=& \bar{v} (k_2) \gamma^\alpha u(k_1) e \frac{-g_{\alpha \beta}}{s} \frac{e m_{R}^2}{f_{R}} \frac{1}{s - m_{R}^2+i m_{R} \Gamma_{R}}\nonumber\\
&&\times \left( -g^{\beta \mu} + \frac{p_0^\beta p_0^\mu}{m_{R}^2} \right) \mathcal{A}^{(i)}_{\mu\nu} \epsilon^\ast_{\psi^\prime}(p_3)^{\nu} \mathcal{F}_{R}(s),
\end{eqnarray}
where $f_{R}$ is a decay constant of an intermediate $R$ and $p_0$ denotes the momentum of the intermediate $R$, which satisfies $p_0^2=(k_1+k_2)^2=s$. $\mathcal{A}_{\mu \nu}$ indicates the decay amplitude of $\psi_R \to \psi(3686) \pi^+ \pi^-$. Here, we also introduce the form factor $\mathcal{F_{R}}$ to balance the otherwise over-increased decay width with an increased phase space, which isgiven by,
\begin{eqnarray}
\mathcal{F}_{R}(s) = e^{-a_{R} \left(\sqrt{s}-m_{R}\right)}.
\end{eqnarray}
where $a_{R}$ is a parameter obtained when fitting the data.

For Fig. \ref{Fig:Mech} (2), the dipion is produced directly by the gluon emission. Hereafter, such a process is named as a direct production process and the decay amplitude is parameterized as \cite{Novikov:1980fa},
\begin{eqnarray}
\mathcal{A}^{(2)}_{\mu\nu} &=& \frac{F_{\rm Dir}}{f_\pi^2} g_{\mu\nu} \Big\{\left[m_{\pi\pi}^2-\kappa_{\rm Dir} (\Delta
M)^2\left(1+\frac{2m^2_\pi}{m_{\pi\pi}^2}\right)\right]_{\mathrm{S-wave}}\nonumber\\
&&+\Big[\frac{3}{2}\kappa_{\rm Dir}\left((\Delta M)^2-m_{\pi\pi}^2\right)
\left(1-\frac{4m_\pi^2}{m_{\pi\pi}^2}\right)\nonumber\\
&&\times\Big(\cos^2\alpha-\frac{1}{3}\Big)\Big]_{\mathrm{D-wave}}\Big\},\label{direct}
\end{eqnarray}
where the terms in the first and second square brackets indicate that the orbital momenta of the dipion are $S$-wave and $D$-wave, respectively.  $\Delta M=m_{R}-m_{\psi^\prime} $ is the mass splitting of the initial $\psi_R$ and final $\psi(3686)$.  $F_{\rm Dir}$ and $\kappa_{\rm Dir}$ are parameters necessary for fitting and $\alpha$ is the angle between $\psi_R$ and $\pi^-$ in the $\pi^+ \pi^-$ rest frame.

As shown in Fig. \ref{Fig:Mech} (3), the dipion is produced through a scalar meson, and the scalar meson as well as $\psi(3686)$ are connected to $\psi_R$ via a charmed meson loop. Such a kind of a meson loop has been proven to be important in the dipion transitions between heavy quarkonia \cite{Chen:2011qx, Chen:2011jp}. Due to the kinetic limit of the involved process, we take the $\sigma$ meson into consideration. Moreover, in the present work, we mainly aim at the ISPE mechanism. Thus, we will not directly estimate the triangle diagrams but parameterize their contributions in the form~\cite{Chen:2013coa},
\begin{eqnarray}
\mathcal{A}^{(3)}_{\mu\nu} &=& \frac{1}{m_{\pi\pi}^2-m_{\sigma}^2 + im_{\sigma} \Gamma_{\sigma}} \Big( f_{\sigma} g_{\mu\nu} +e^{i \varphi_{\sigma}} g_{\sigma} p_{0\nu} p_{3 \mu} \Big),\nonumber\\\label{LoopS}
\end{eqnarray}
where $f_{\sigma}$ and $g_\sigma$ are the $S$ and $D$ wave coupling constants between $\psi_R$ and $\psi(3686) \mathcal{S}$, respectively, and $\varphi_\sigma$ is the phase angle between the $S$ and $D$ wave couplings. We should notice that the mass and width of the $\sigma$ meson are comparable, and thus, we adopt a momentum dependent width, which is,
\begin{eqnarray}
\Gamma_\sigma (m_{\pi\pi}) = \Gamma_{\sigma} \frac{m_{\sigma}}{m_{\pi \pi}} \frac{\left| \vec{P}(m_{\pi\pi}) \right|}{\left| \vec{P}(m_\sigma) \right|}
\end{eqnarray}
where $\left| \vec{P}(m_{\pi\pi})\right|=\sqrt{m_{\pi\pi}^2/4-m_\pi^2}$ is the pion momentum in the dipion rest frame, while $|\vec{P}(m_{\sigma})|$ is the pion momentum with the on-shell $\sigma$ meson.

As for the ISPE mechanism, we directly estimate the triangle diagrams. To fit with the experimental data, we reduce the amplitudes of the ISPE mechanism in the form,
\begin{eqnarray}
\mathcal{A}^{(4)}_{\mu\nu} &=& C_{00} g_{\mu\nu} + C_{12} p_{1\mu}p_{2\nu} + C_{13} p_{1\mu}p_{3\nu}\nonumber\\
&&+ C_{22} p_{2\mu}p_{2\nu} + C_{23} p_{2\mu}p_{3\nu}\nonumber\\
&&+ C_{32} p_{3\mu}p_{2\nu} + C_{33} p_{3\mu}p_{3\nu},\label{eqs:ISPE-Parametrization}
\end{eqnarray}
where the coefficients $C_{ij}, \ \{(i,j)=(0,1,2,3)\}$ are obtained by the loop integrals of the amplitudes in Appendix \ref{Sec:App-Amp}.

\begin{figure}[htbp]
\begin{center}
\scalebox{0.330}{\includegraphics{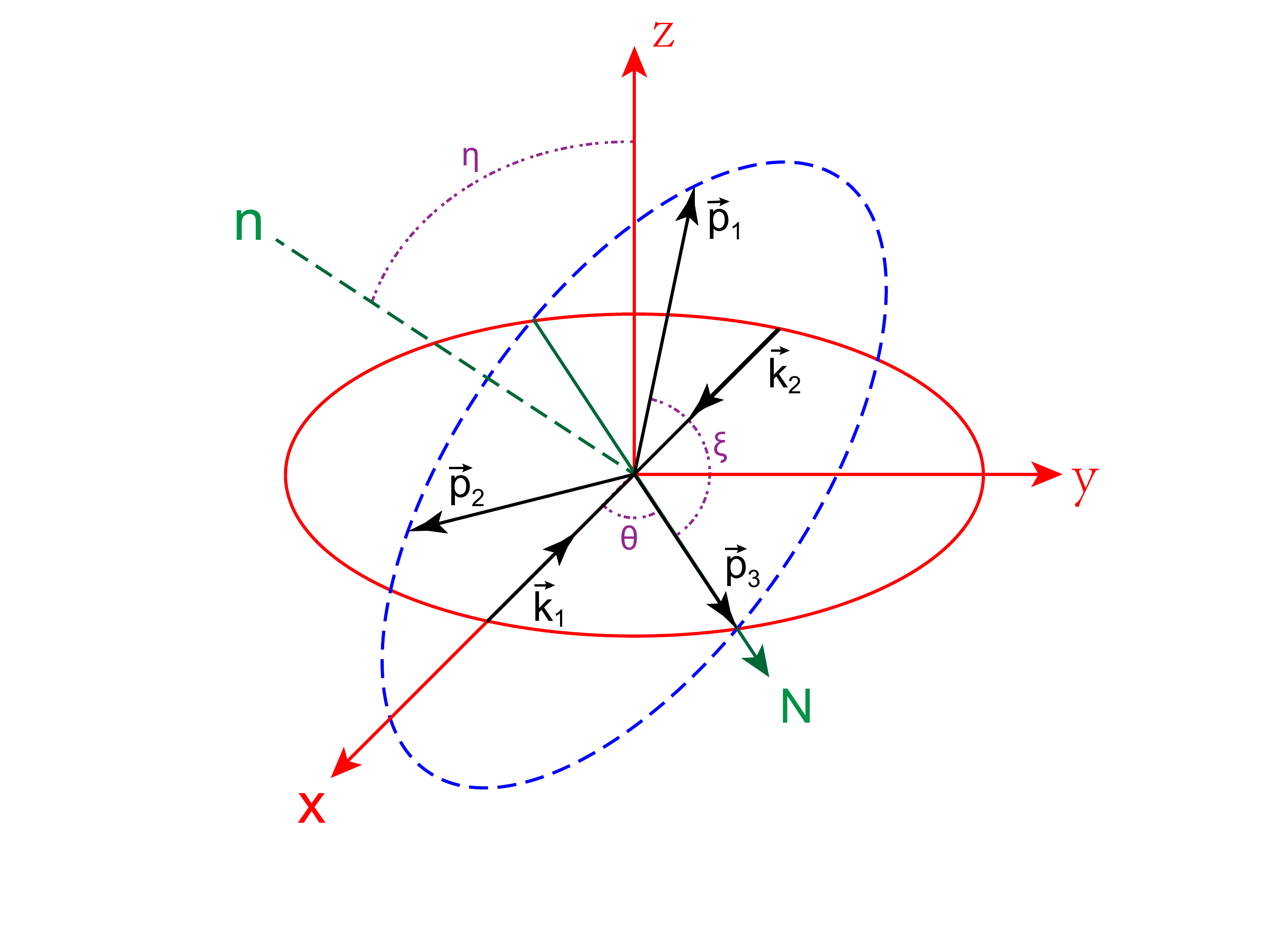}}
\caption{(Color online). The definitions of angles $\theta$ and $\eta$ that appear in Eq. (\ref{eqs:phase}). In this figure, we also give the relative positions of $\vec{k}_1$, $\vec{k}_2$, and $\vec{p}_i~(i=1,2,3)$. Since $p_0=k_1+k_2=(\sqrt{s},0,0,0)$, $\vec{p}_0$ lies on the origin of the $xyz$ frame.}\label{fig:Euler-angle}
\end{center}
\end{figure}

With above preparations, we get the total amplitude, which could be the coherent sum of the amplitudes corresponding to different mechanisms, and reads as
\begin{eqnarray}
\mathcal{M}^{\mathrm{Tot}}=\mathcal{M}_{\rm NoR}+\sum_{R,i=2}^{i=4} e^{i \phi_{R}^{(i)}}  \mathcal{M}_{R}^{(i)}
\label{eqs:total}
\end{eqnarray}
with $\phi_{R}^{(i)}$ being the phase angles in front of different amplitudes for each resonance. For convenience, we use $\phi_{\rm Dir}$, $\phi_\sigma$ and $\phi_{\rm ISPE}$ to refer to these phase angles in the amplitudes of  direct production process, $\sigma$ production process and ISPE processes, respectively. The differential cross section is \cite{Xie:2015zga}
\begin{eqnarray}
d\sigma_{e^+e^- \to \psi(3686) \pi^+ \pi^-}=\frac{(2\pi)^4 \sum \left|\overline{\mathcal{M}^{\rm Tot}} \right|^2}{4 (k_1\cdot k_2)}d\Phi_3,\label{eqs:cross}
\end{eqnarray}
where $\sum \left|\overline{\mathcal{M}^{\rm Tot}} \right|^2$ indicates an average over the spin of initial electron and positron and a sum over the spin of the final state $\psi(3686)$. $d\Phi_3$ is the phase space factor, which is expressed as
\begin{align}
d\Phi_3=\frac{1}{32(2\pi)^8 s}   dm_{12}^2 dm_{23}^2 d(\mathrm{cos}\theta)d\eta ,
\label{eqs:phase}
\end{align}
where $m_{23}$ corresponds to $m_{\psi^\prime \pi^+}$, $m_{12}$ is $m_{\pi^+\pi^-}$, and the definitions of angles $\theta$ and $\eta$ are shown in Fig. \ref{fig:Euler-angle}.

\section{Numerical Results and discussion}
\label{Sec:Num}
In the differential cross section in Eq. (\ref{eqs:cross}), there are two free parameters, $g_{\rm NoR}$ and $a_{\rm NoR}$,  in the nonresonance amplitude. As for the intermediate charmonium contributions, there are 12 additional free parameters for each resonance in Table \ref{tab:psi-parameter}, which are
\\

{\it
\begin{tabular}{p{50pt}p{5pt}p{155pt}}
$a_R$ &:& a parameter in the resonance form factor,\\
$\phi_{\rm Dir}$ &:& a phase angle of the direct production amplitude,\\
$F_{\rm Dir},\ \kappa_{\rm Dir}$ &:& parameters in the direct production amplitude,\\
$\phi_\sigma$ &:& a phase angle of the $\sigma$ meson production amplitude,\\
$f_\sigma, \ g_\sigma$ &: & $S$ and $D$ wave coupling constants in the $\sigma$ meson production amplitude,\\
$\varphi_\sigma$ &:& a relative phase angle between $S$ and $D$ wave terms in the $\sigma$ meson production amplitude,\\
$\phi_{\rm ISPE}$&: &a phase angle of the $\mathrm{ISPE}$ amplitude,\\
$g_{R D^{(\ast)} \bar{D}^{(\ast)} \pi}$ &: & coupling constants of the four-particle interactions in the $\mathrm{ISPE}$ mechanism.
\end{tabular}
}

\begin{center}
    \begin{figure*}[htbp]
    \begin{tabular}{cccc}
        \scalebox{0.145}{\includegraphics{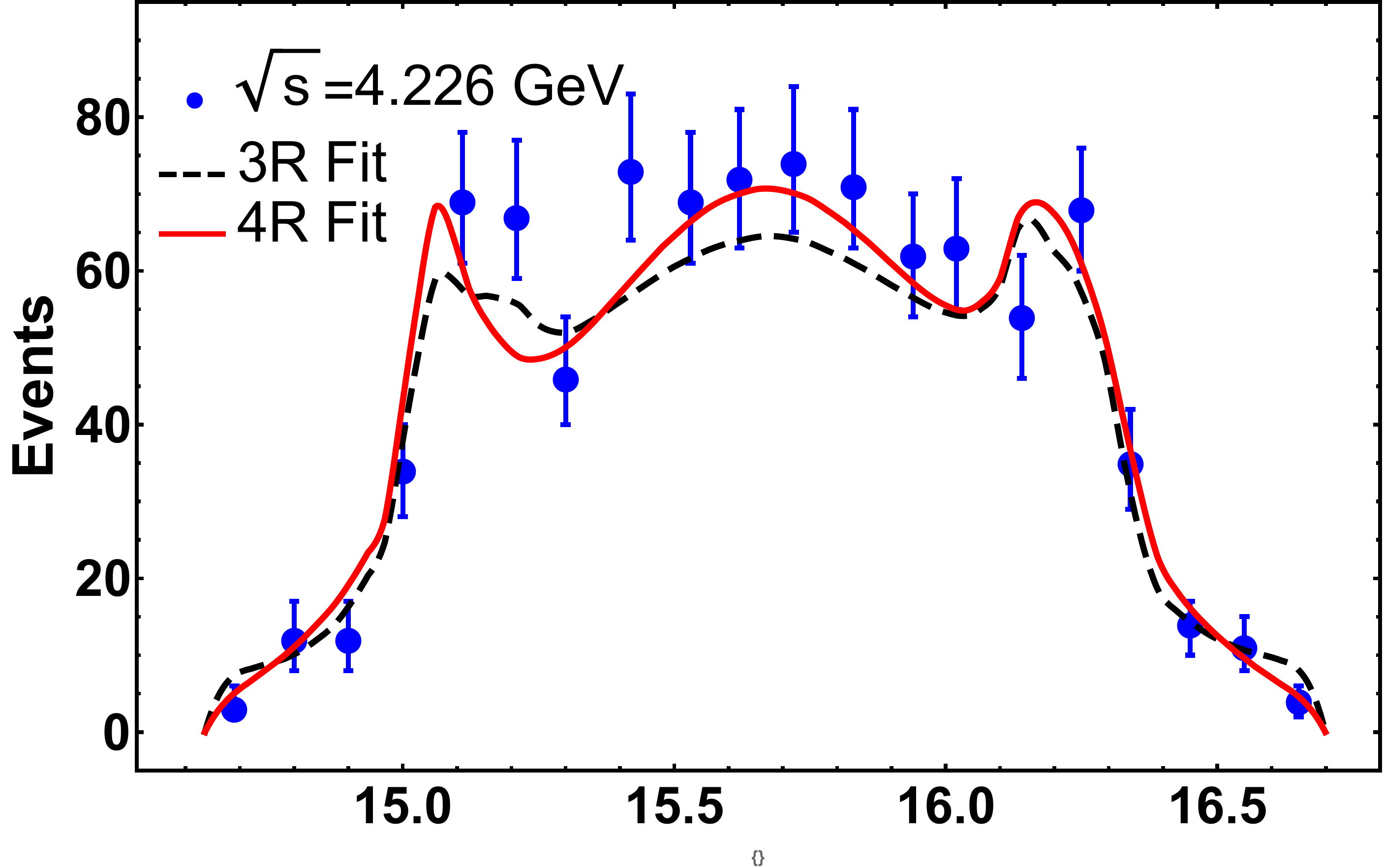}} & \scalebox{0.140}{\includegraphics{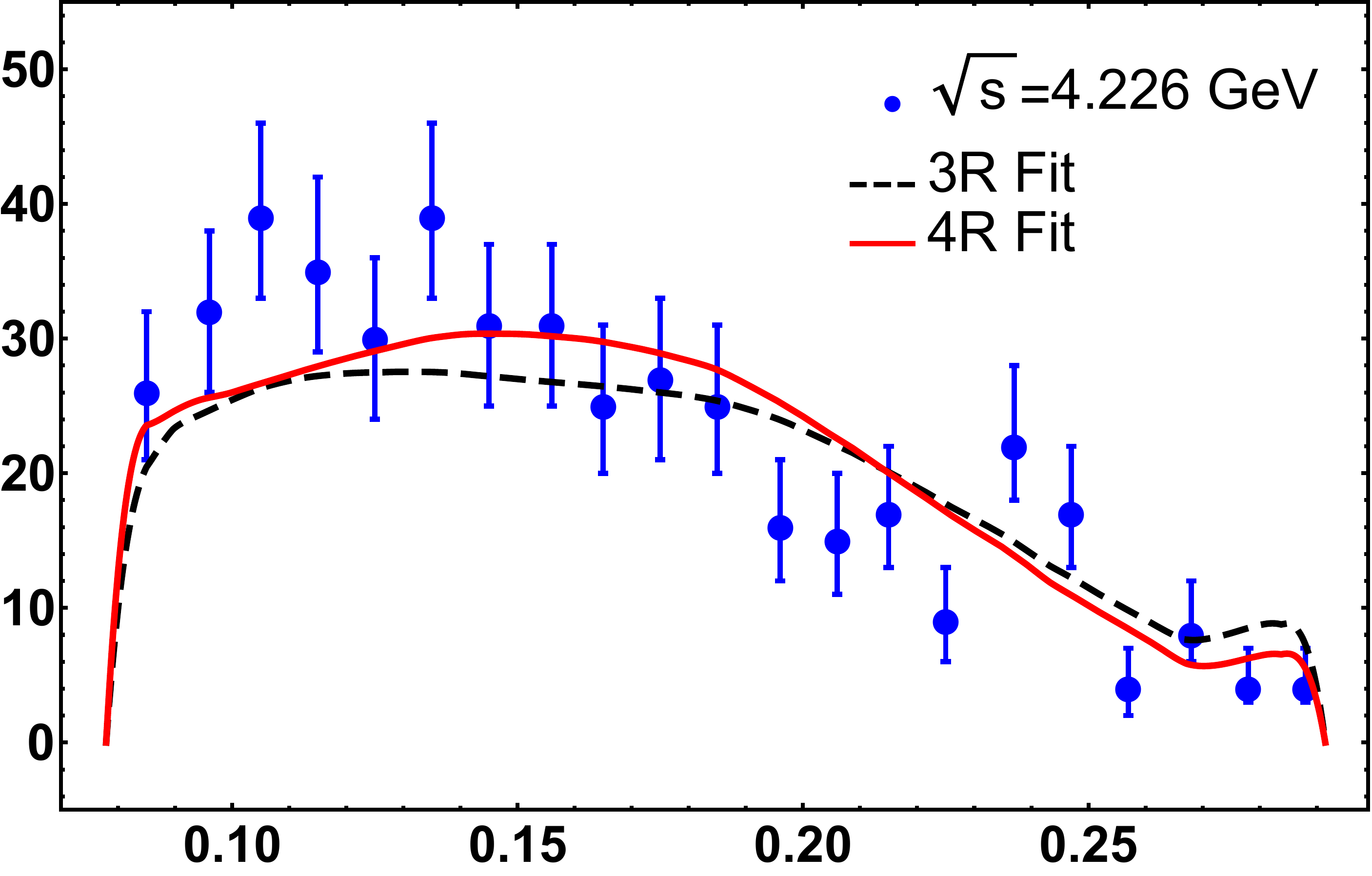}} & \scalebox{0.140}{\includegraphics{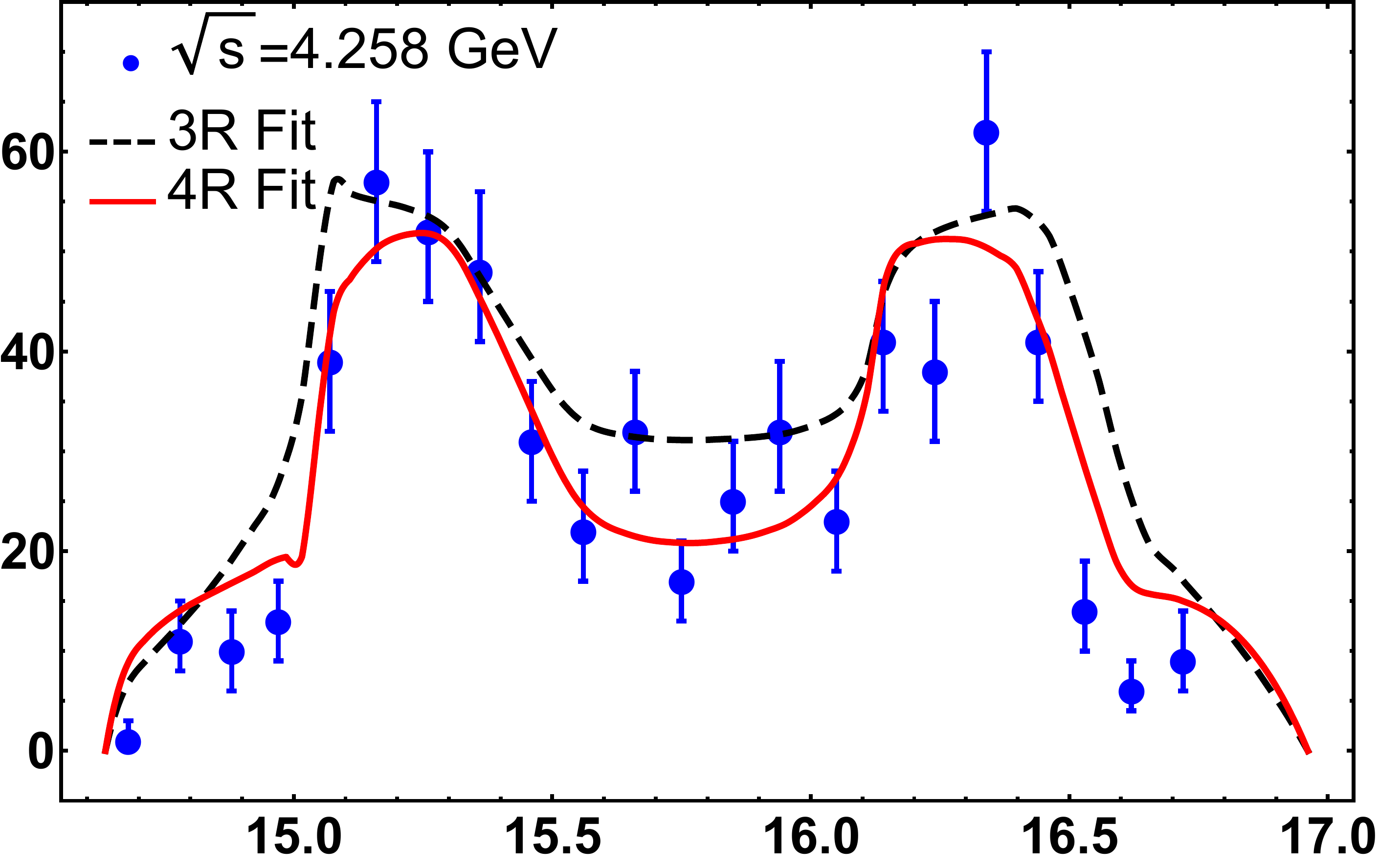}} & \scalebox{0.140}{\includegraphics{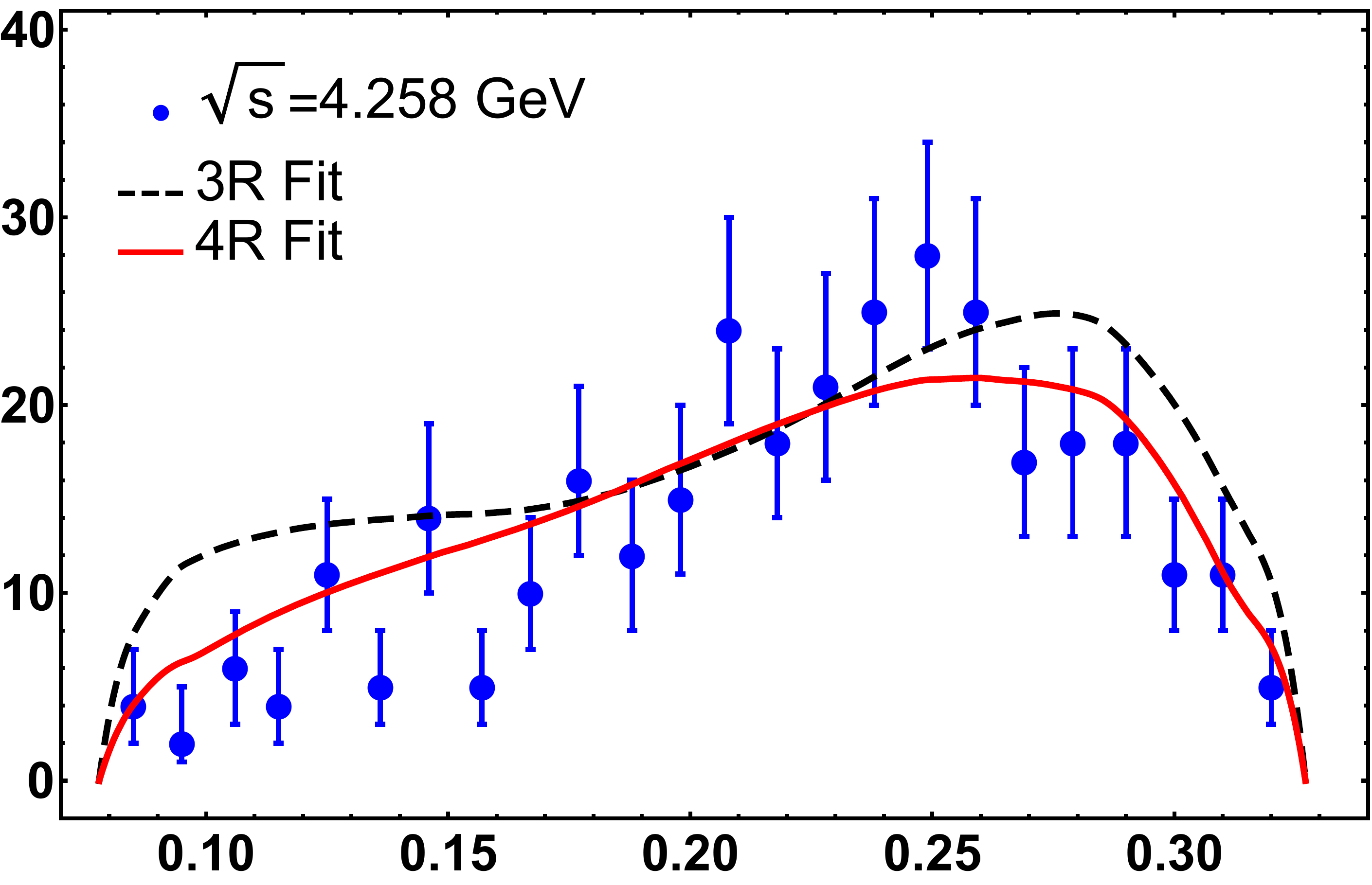}}\\
        \scalebox{0.145}{\includegraphics{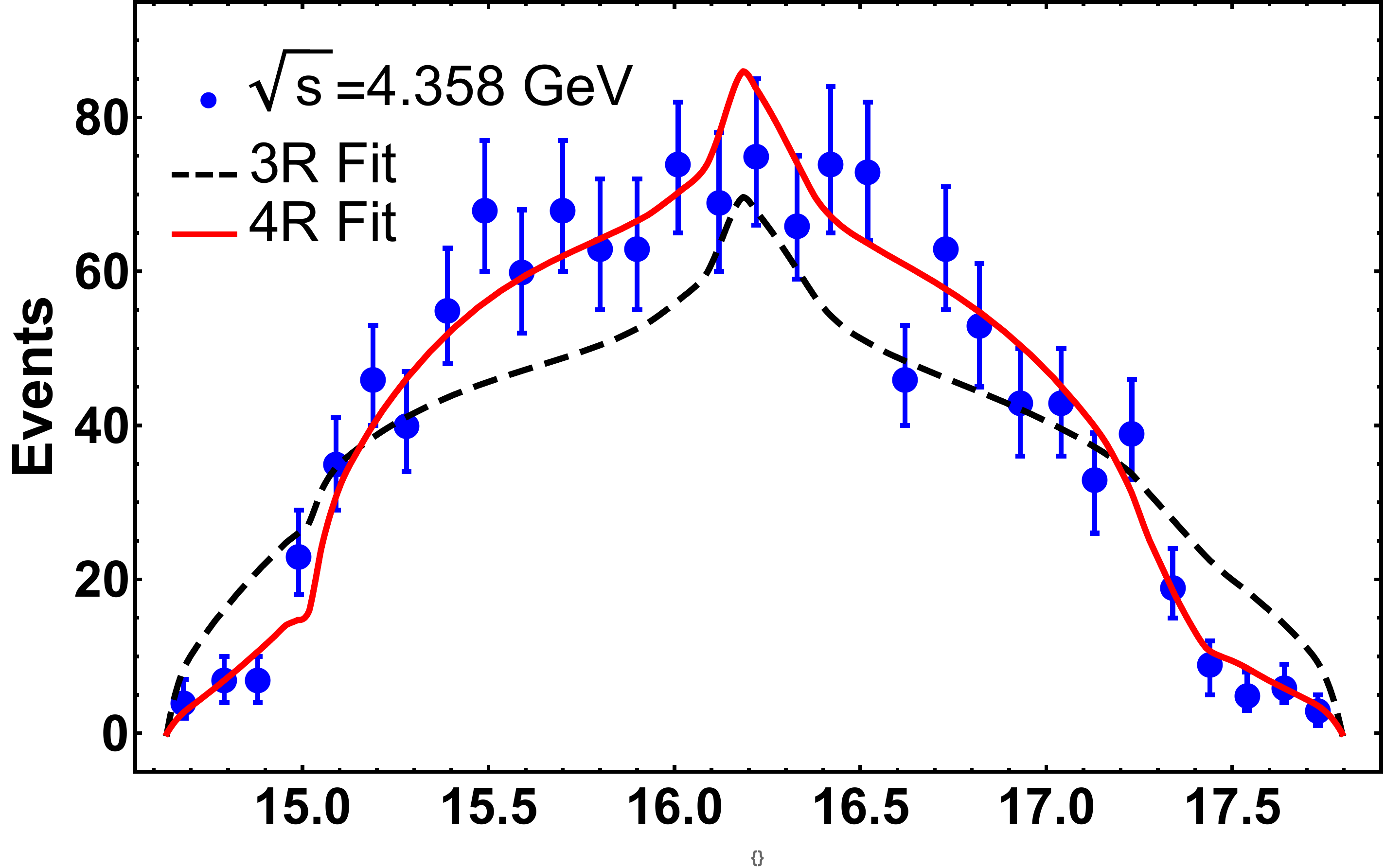}} & \scalebox{0.140}{\includegraphics{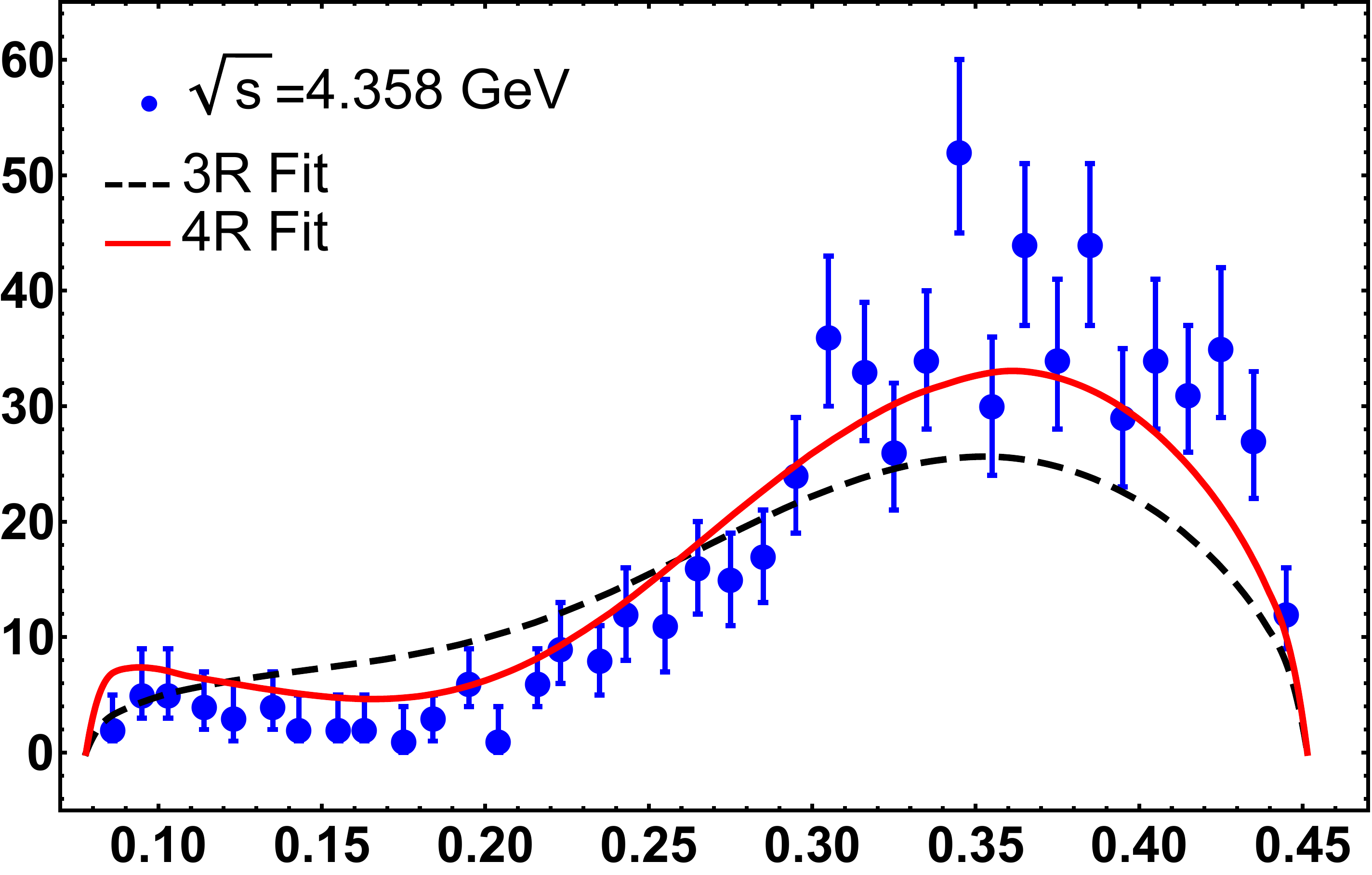}} & \scalebox{0.140}{\includegraphics{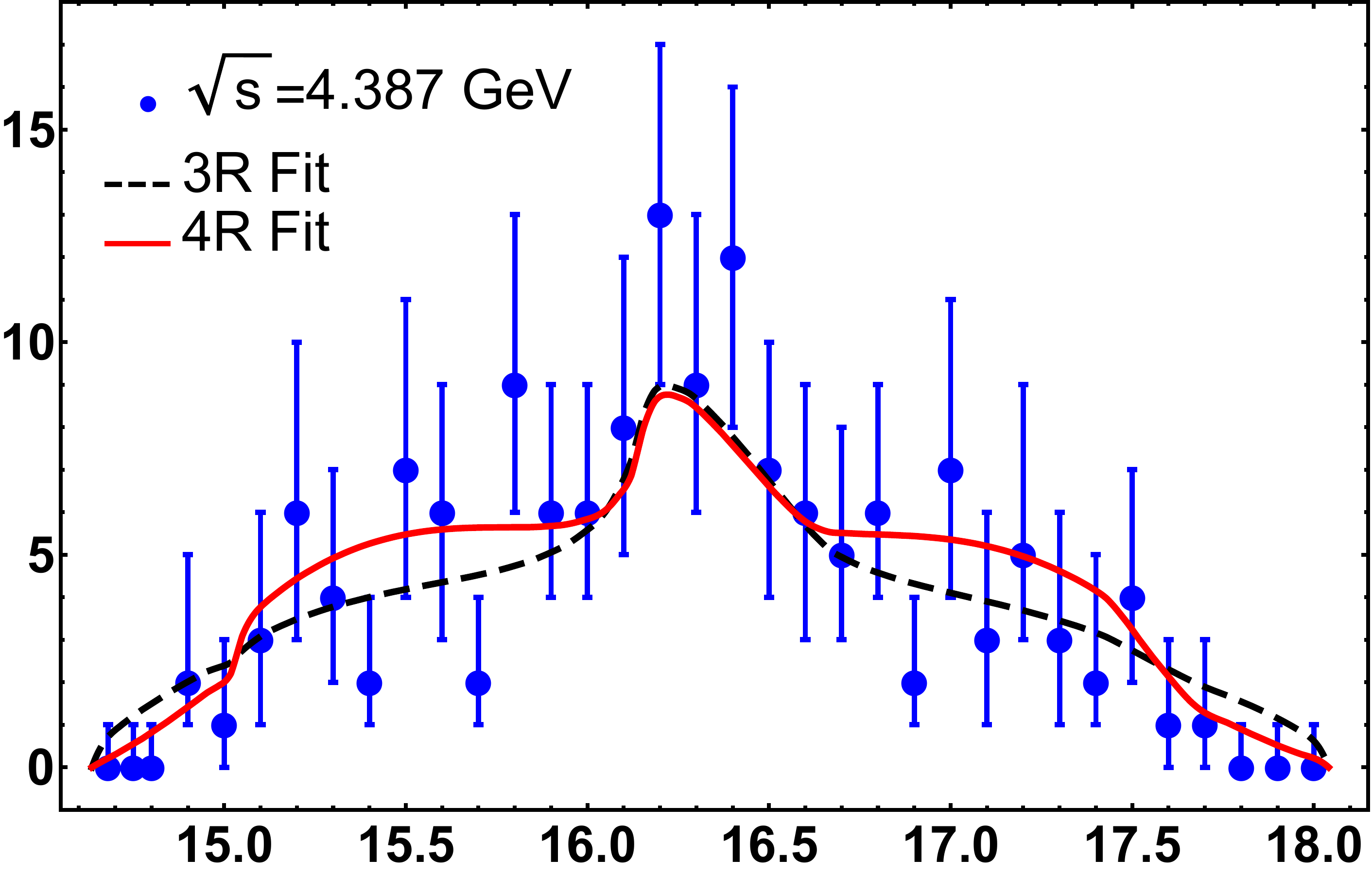}} & \scalebox{0.140}{\includegraphics{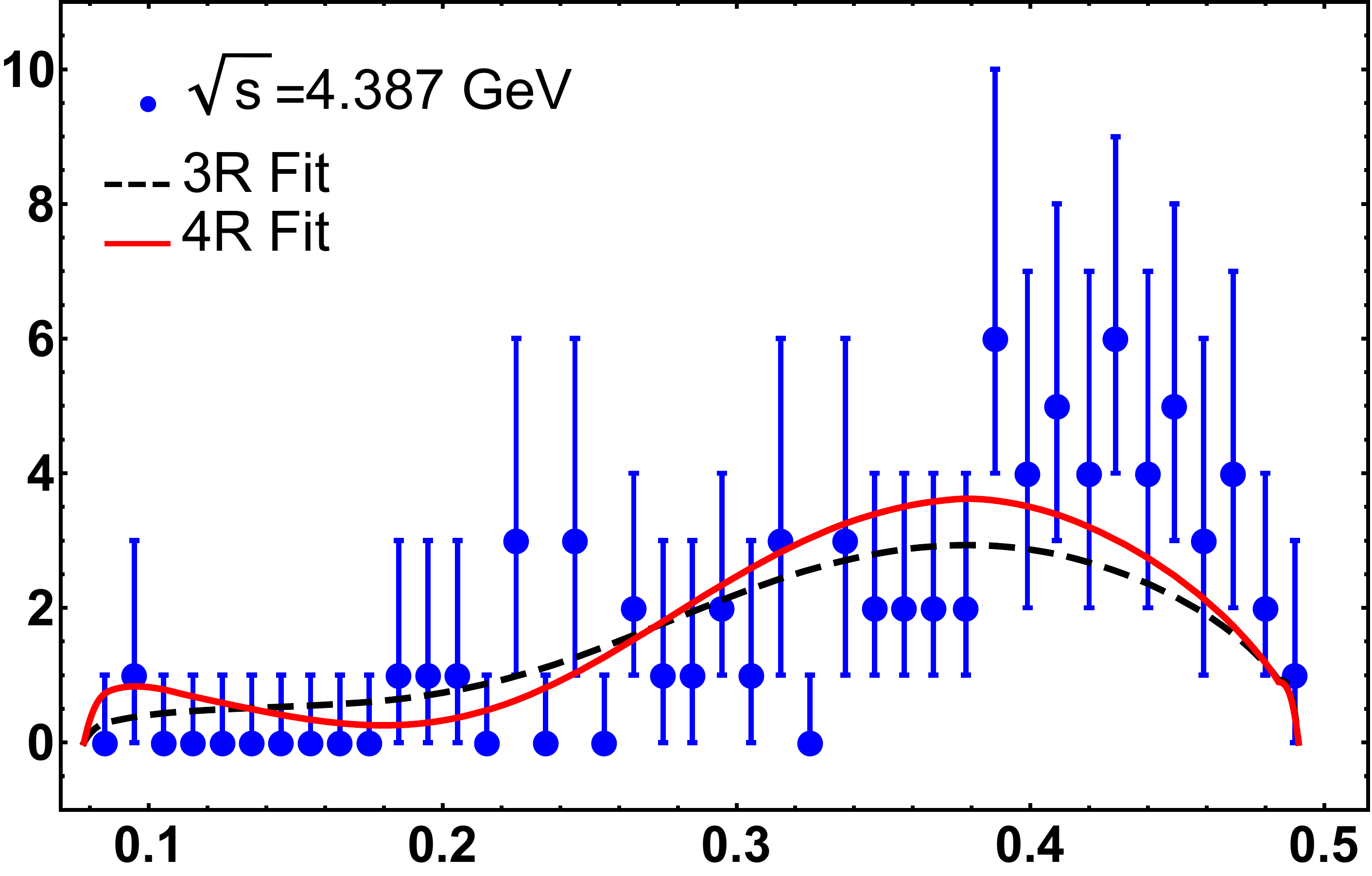}}\\
        \scalebox{0.145}{\includegraphics{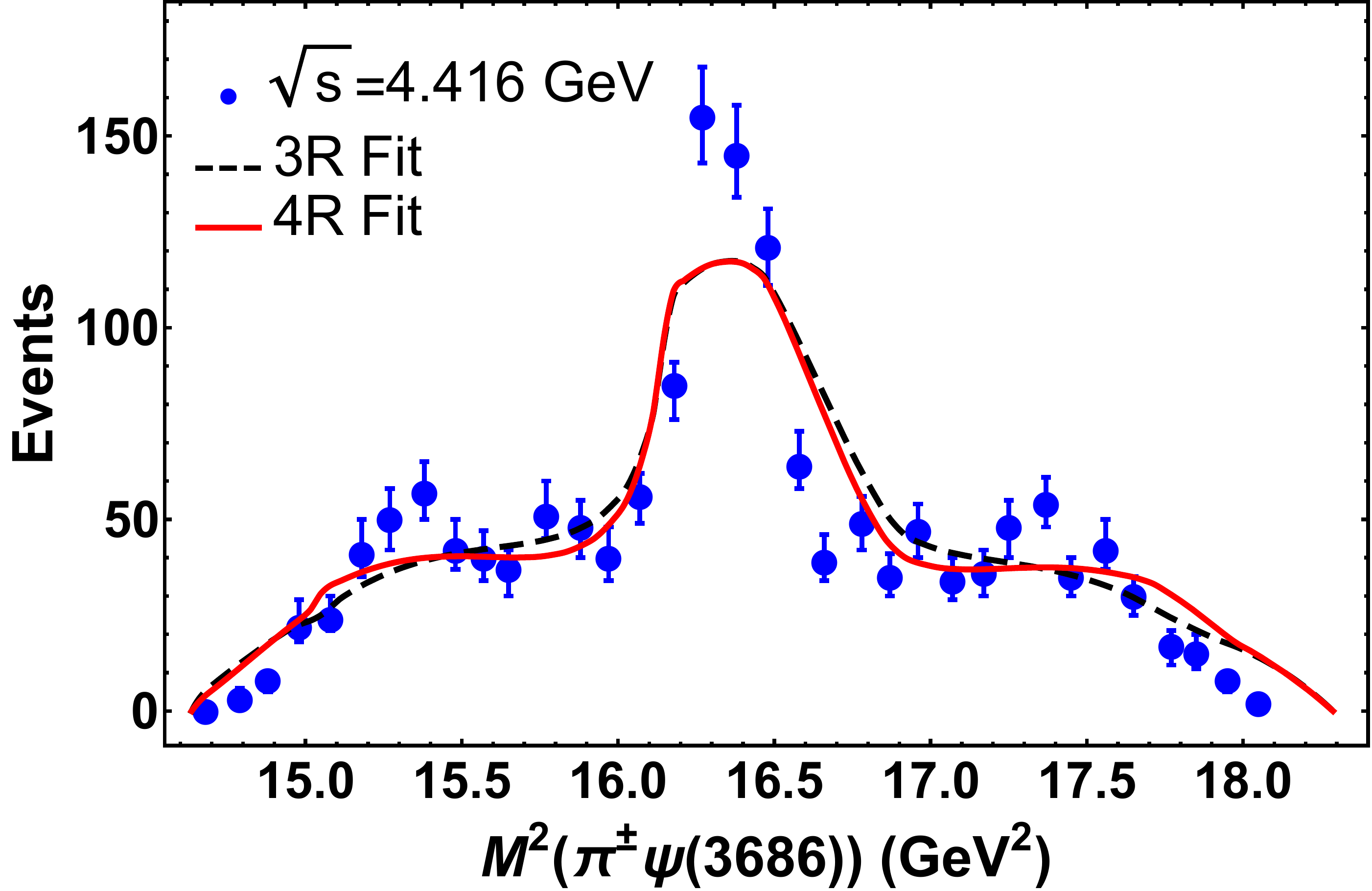}} & \scalebox{0.140}{\includegraphics{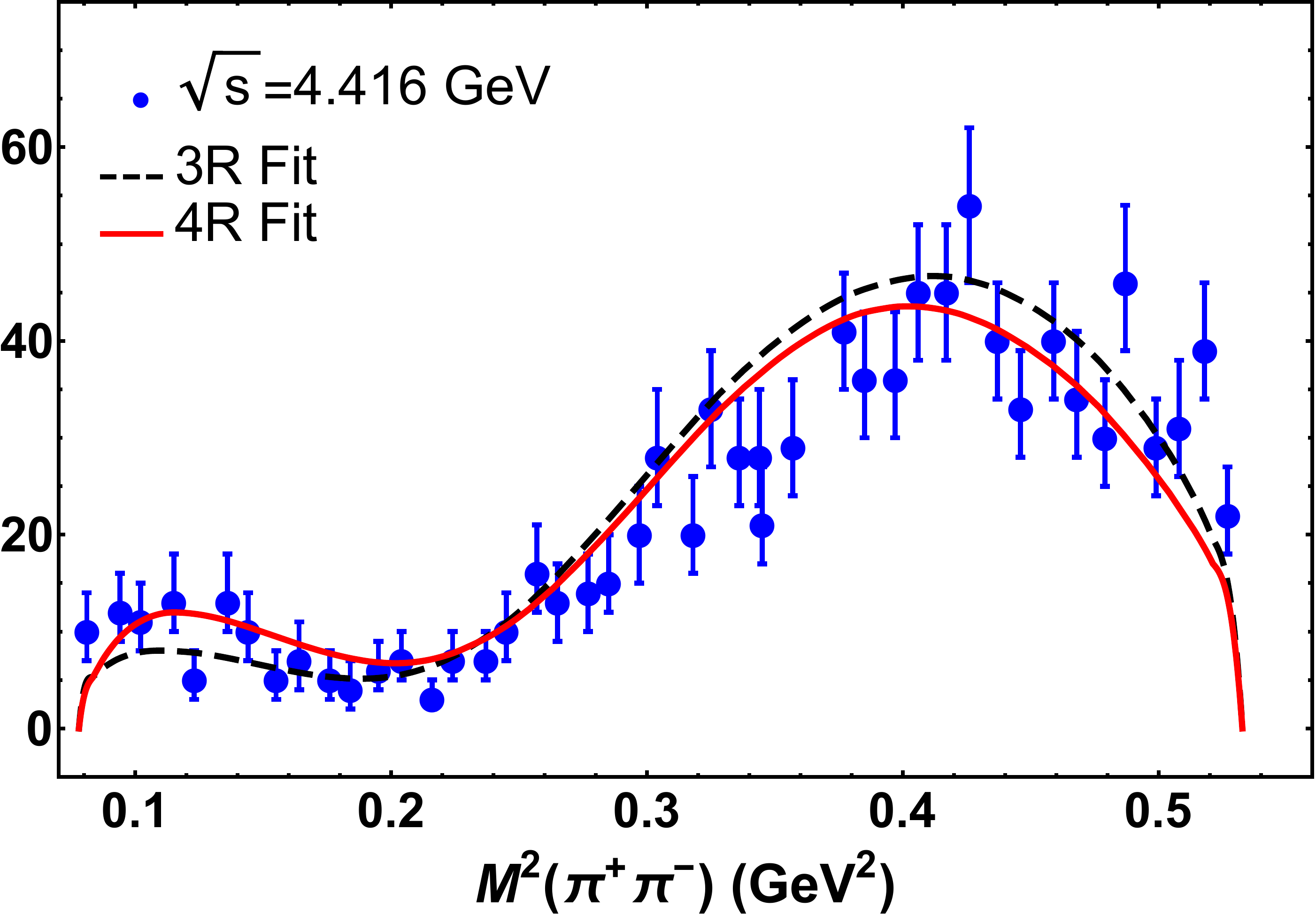}} & \scalebox{0.140}{\includegraphics{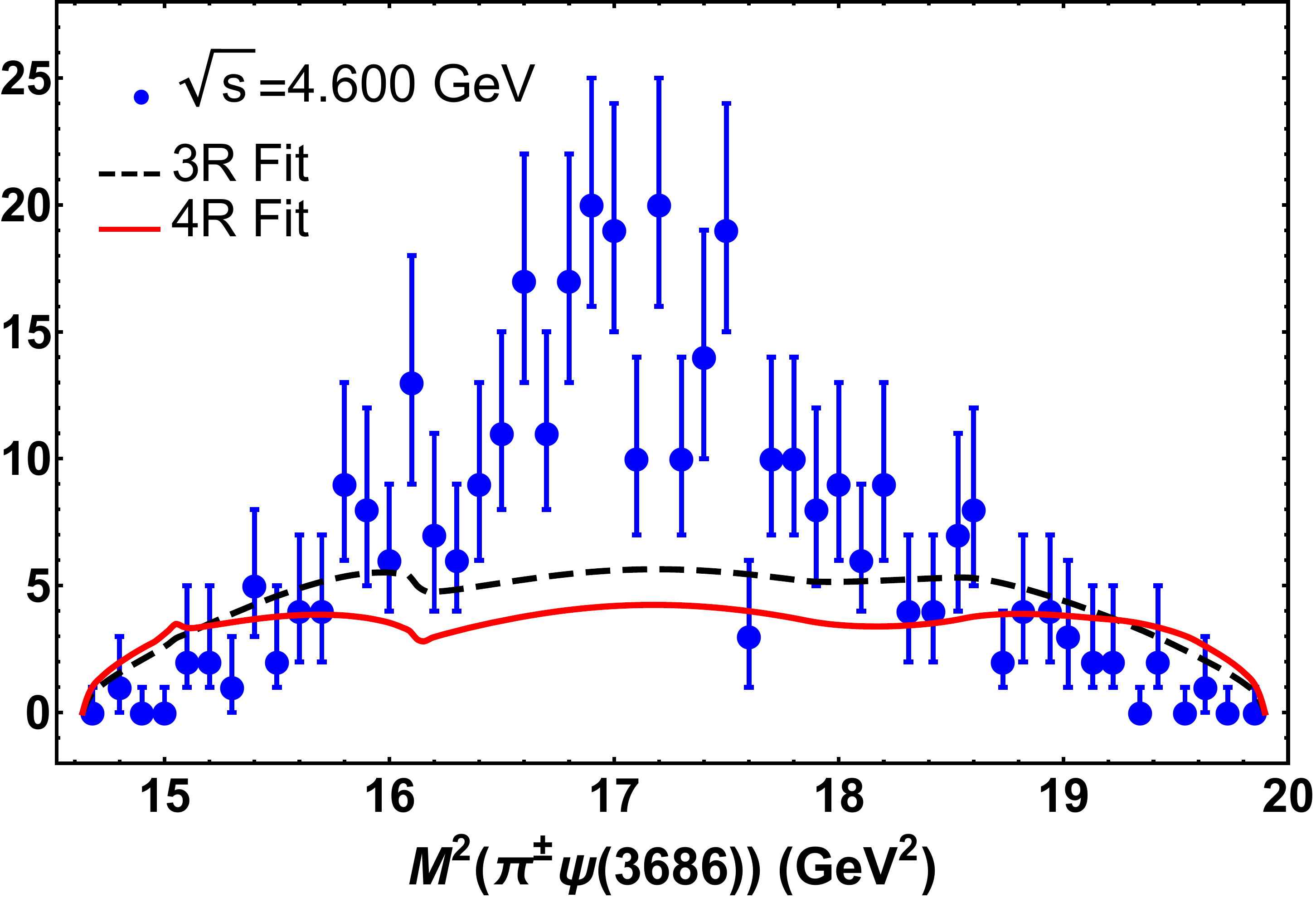}} & \scalebox{0.140}{\includegraphics{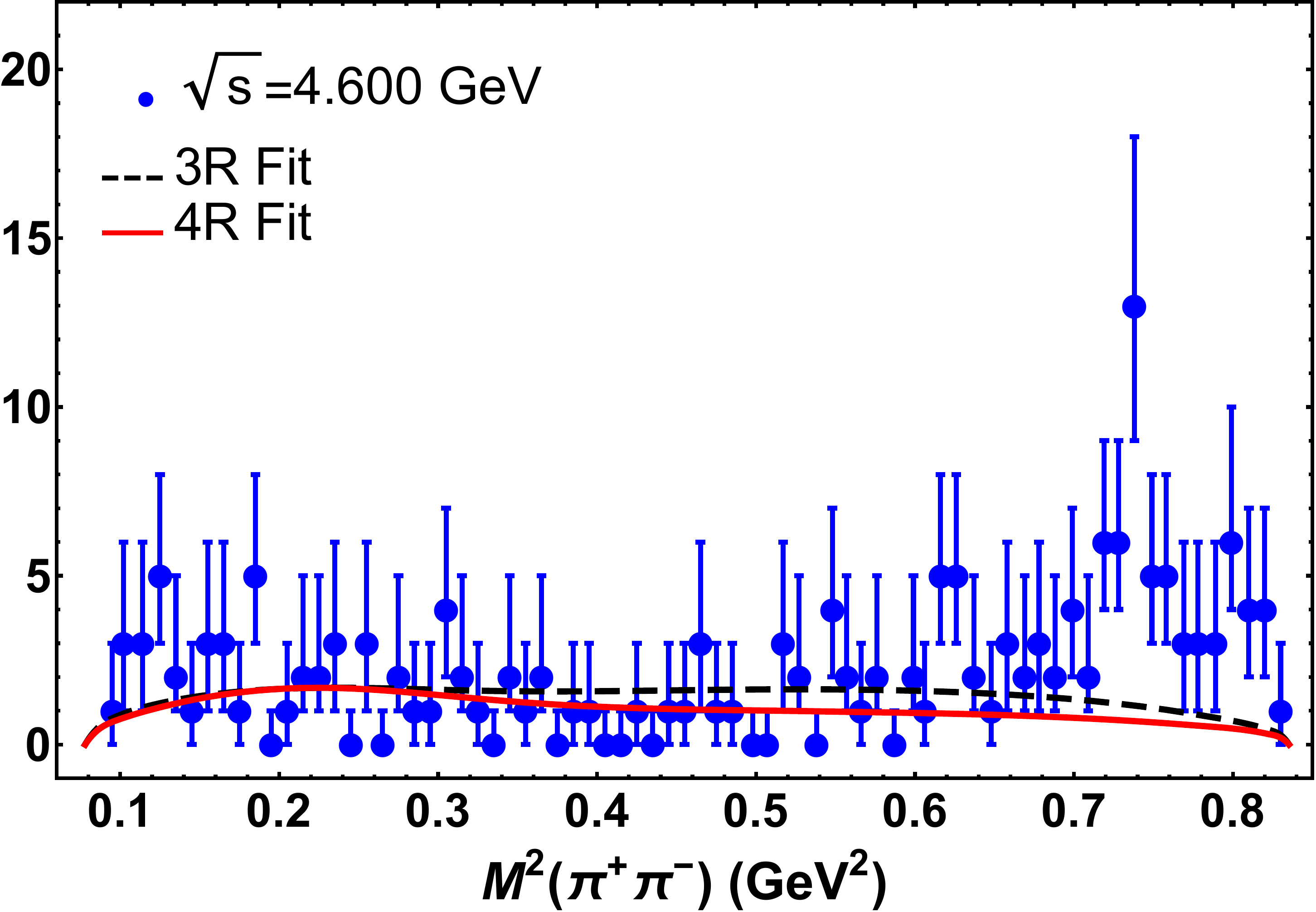}}
    \end{tabular}
	\caption{(Color online). Our combined fit for the $\psi(2S)\pi^\pm$ and $\pi^+\pi^-$ invariant mass spectra lying on 6 different center-of-mass energies which are reported by BESIII in Ref. \cite{Ablikim:2017oaf}. Here, the black dashed and red solid curves are the fitted results under three-charmonium and four-charmonium scenarios, respectively.}\label{fig:ims}
	\end{figure*}
\end{center}

In the following, we need to fit our results with the experimental data of $e^+e^- \to \psi(3686) \pi^+ \pi^-$, which include the measured total cross section, the $\psi(3686) \pi^\pm$, and $\pi^+\pi^-$ invariant mass spectra released in Ref. \cite{Ablikim:2017oaf}. For the $e^+e^- \to \psi(3686) \pi^+ \pi^-$ process, there exist contributions from intermediate vector charmonia as shown in Fig. \ref{Fig:Mech} (2)-(4). Thus, it is a key point how to select intermediate vector charmonia. In the present work, we adopt two scenarios, which will be addressed in Sec. \ref{three} and Sec. \ref{four}.

\subsection{Three-charmonium scenario to $e^+e^- \to \psi(3686) \pi^+ \pi^-$}\label{three}

When checking the vector charmonium states located in 4.1-4.6 GeV listed in particle data book \cite{Tanabashi:2018oca}, there exist only two observed charmonia, $\psi(4160)$ and $\psi(4415)$. Besides, we must mention that there is another well-established charmonium-like state $Y(4220)$ in this energy range. This state has been observed experimentally in the processes of $e^+ e^- \to \chi_{c0} \omega$
\cite{Ablikim:2014qwy}, $e^+ e^- \to \pi^+ \pi^- h_c$ \cite{{BESIII:2016adj}}, $e^+ e^- \to \pi^+ \pi^- J/\psi$ \cite{Ablikim:2016qzw}, $e^+ e^- \to \pi^+ D^0 D^{\ast -}$ \cite{Ablikim:2018vxx}, and $e^+ e^- \to \pi^+ \pi^- \psi(3686)$ \cite{Ablikim:2017oaf}. The mass of this state is consistent with the predictions of a missing higher charmonium in the $J/\psi$ family (see the discussion in Refs. \cite{He:2014xna,Wang:2019,Chen:2014sra}). Thus, $Y(4220)$ is replaced by the renamed $\psi(4220)$ in the following discussion.

In the present section, we first try to fit the experimental data by considering $\psi(4160)$, $\psi(4220)$, and $\psi(4415)$ as intermediate state contributions to $e^+ e^- \to \pi^+ \pi^- \psi(3686)$. Their resonance parameters 
are collected in Table \ref{tab:psi-parameter}.

\renewcommand{\arraystretch}{1.5}
\begin{table}[htpb]
\centering \caption{The resonance parameters and dilepton decay widths of the involved states \cite{Tanabashi:2018oca}. The dilepton decay widths of $\psi(4220)$ and $\psi(4380)$ are taken from Ref. \cite{Wang:2019} }
\label{tab:psi-parameter}
\begin{tabular}{p{40pt}<\centering p{60pt}<\centering p{60pt}<\centering p{60 pt}<\centering}
\toprule[1pt]
Meson  & Mass (GeV) & Width (MeV) & $\Gamma_{e^+ e^-}$ (keV)\\
\midrule[0.6pt] %
$\psi(4160)$ & 4.191 & 70 & 0.48\\
$\psi(4415)$ & 4.421 & 62 & 0.58\\
$\psi(4220)$ & 4.218 & 59 & 0.29\\
$\psi(4380)$   & 4.384  & 84  & 0.26\\
\bottomrule[1pt]
\end{tabular}
\end{table}

\begin{table}[htpb]
\centering \caption{The fitted values of the parameters in three-charmonium scenario.}\label{tab:fit-parameter}
\begin{tabular}{p{3cm}<\centering p{1.5cm}<\centering p{1.5cm}<\centering p{1.5cm}<\centering }
\toprule[1pt]
Prameters & $\psi(4160)$ & $\psi(4220)$ & $\psi(4415)$\\
\midrule[1pt]
$a_{R}$ (GeV$^{-1}$) & 6.0 & 5.0 & 4.0\\
$\phi_{\rm Dir}$ & -1.631 & -1.506 & 1.762\\
$F_{\rm Dir}$  & -3.477 & 5.881 & 0.982\\
$\kappa_{Dir}$  & -1.829 & -0.025 & -0.918\\
$\phi_{\sigma}$ & 1.130 & 2.009 & 2.241\\
$f_{\sigma}$ (GeV$^2$) & 7.562 & 5.785 & -3.971\\
$g_{\sigma}$ & 77.498 & 21.803 & -1.920\\
$\varphi_{\sigma}$ & 1.186 & -2.426 & -1.923\\
$\phi_{\rm ISPE}$ & -1.320 & 2.980 & -2.813\\
$g_{ R D \bar{D} \pi}$ (GeV$^{-3}$) & 1.717 & 1.574 & 0.002\\
$g_{R D^\ast \bar{D} \pi}$  & -1.834 & -0.670 & 0.025\\
$g_{R D^\ast \bar{D}^\ast \pi}$ (GeV$^{-1}$) & 0.120 & -0.096 & 0.105\\
\midrule[1pt]
$g_{\rm NoR}$=18.940 GeV &  \multicolumn{3}{c}{$a_{\rm NoR}$=4.702 GeV$^{-2}$}\\
\midrule[1pt]
\multicolumn{4}{c}{$\chi^2/d.o.f.=3.704$}\\
\bottomrule[1pt]
\end{tabular}
\end{table}

After integrating over $m_{ \pi^\pm\psi(3686)}$ or $m_{\pi^+ \pi^-}$ in Eq. (\ref{eqs:cross}) at each fixed $\sqrt{s}$, one gets the differential cross sections of $e^+ e^- \to \pi^+ \pi^- \psi(3686)$ for each invariant mass squared distribution. The cross section depending on $\sqrt{s}$ can be obtained by integrating over both $m_{ \psi(3686)\pi^\pm }$ and $m_{\pi^+ \pi^-}$ in Eq. (\ref{eqs:cross}). Here, we perform a combined fit of the invariant mass squared dependent cross sections to the experimental data of the differential cross sections at a fixed $\sqrt{s}$. In the three-resonance scenario, there are 38 parameters, which should be determined by the present fit. Our fitted values for all the parameters are presented in Table \ref{tab:fit-parameter}. With these parameters, $\chi^2/{\rm d.o.f}$ is estimated to be $3.704$.

The fitted differential cross sections for $e^+ e^- \to \pi^+ \pi^- \psi(3686)$ at a fixed $\sqrt{s}$ in three-resonance scenario are presented in Fig. \ref{fig:ims} (black dashed curves), where $\sqrt{s}$ is fixed at $4.226$, $4.258$, $4.358$, $4.387$, $4.416$, and $4.600$ GeV, respectively. Most of the differential cross sections can be reproduced. In particular, both $\pi^\pm \psi(3686)$ and $\pi^+\pi^-$ invariant mass spectra are well reproduced for $\sqrt{s}=4.226$ GeV. As for $\sqrt{s}=4.258$ GeV, the $\pi^\pm \psi(3686)$ invariant mass spectrum is a bit different from the experimental data. In the vicinity of $m^2_{\pi^\pm \psi(3686)}=15\ {\rm GeV}^2$, the experimental data show a jump, but such a feature is not reproduced in three-charmonium scenario. On the other hand, the fitted curve of $m_{\pi^+ \pi^-}$ distribution is slightly larger than the experimental data in the low and high $m_{\pi^+ \pi^-}$ energy range. Similar to the case for the fit to the data at 4.358 GeV, the fitted curve of the $m^2_{\pi^+ \psi(3686)}$ distribution is a bit lower on the whole than the experimental data, while for $m^2_{\pi^+ \pi^-}$ distribution, the fitted curve is a little bit larger than the experimental data in the lower $m_{\pi^+ \pi^-}$ energy range. For the cases of $\sqrt{s}=4.387$ GeV and $\sqrt{s}=4.416$ GeV, the fitted curves for both $m^2_{\pi^\pm \psi(3686)}$ and $m^2_{\pi^+ \pi^-}$ well reproduce the experimental data. As for the data at $4.600$ GeV, the fitted curve obtained under three-charmonium scenario is lower than the experimental data for the $m^2_{\pi^\pm \psi(3686)}$ invariant mass distribution, while for $m^2_{\pi^+ \pi^-}$ distribution, since the errors of experimental data are very large, thus, the fitted curve is roughly consistent with the data.

\begin{center}
	\begin{figure}[htbp]
		\scalebox{0.24}{\includegraphics{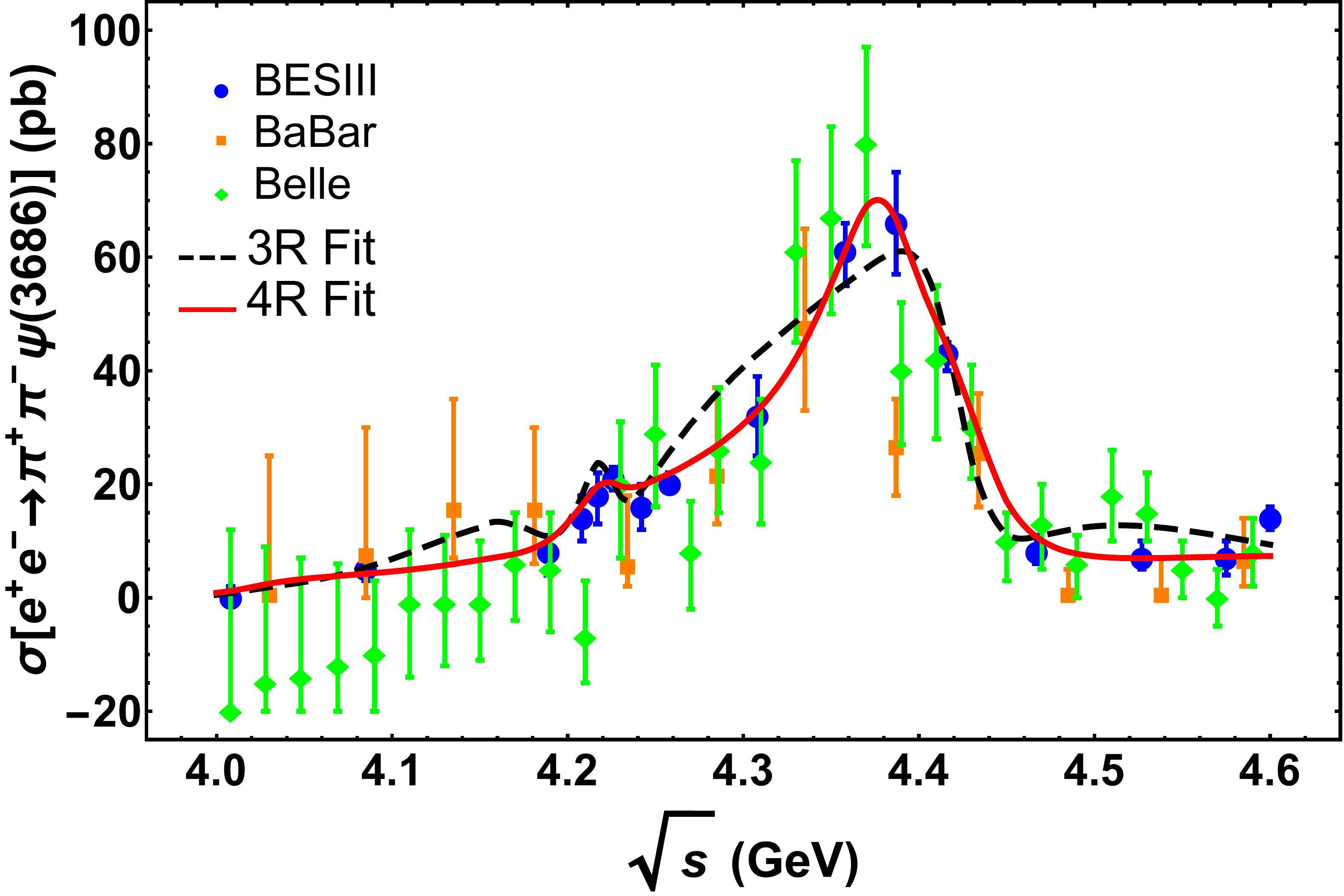}}
		\caption{(Color online). A combined fit of the cross section for $e^+e^-\to\psi(2S)\pi^+\pi^-$ given by BESIII in Ref. \cite{Ablikim:2017oaf}.}
		\label{fig:cs}
	\end{figure}
\end{center}

\begin{center}
	\begin{figure}[htbp]
		\scalebox{0.225}{\includegraphics{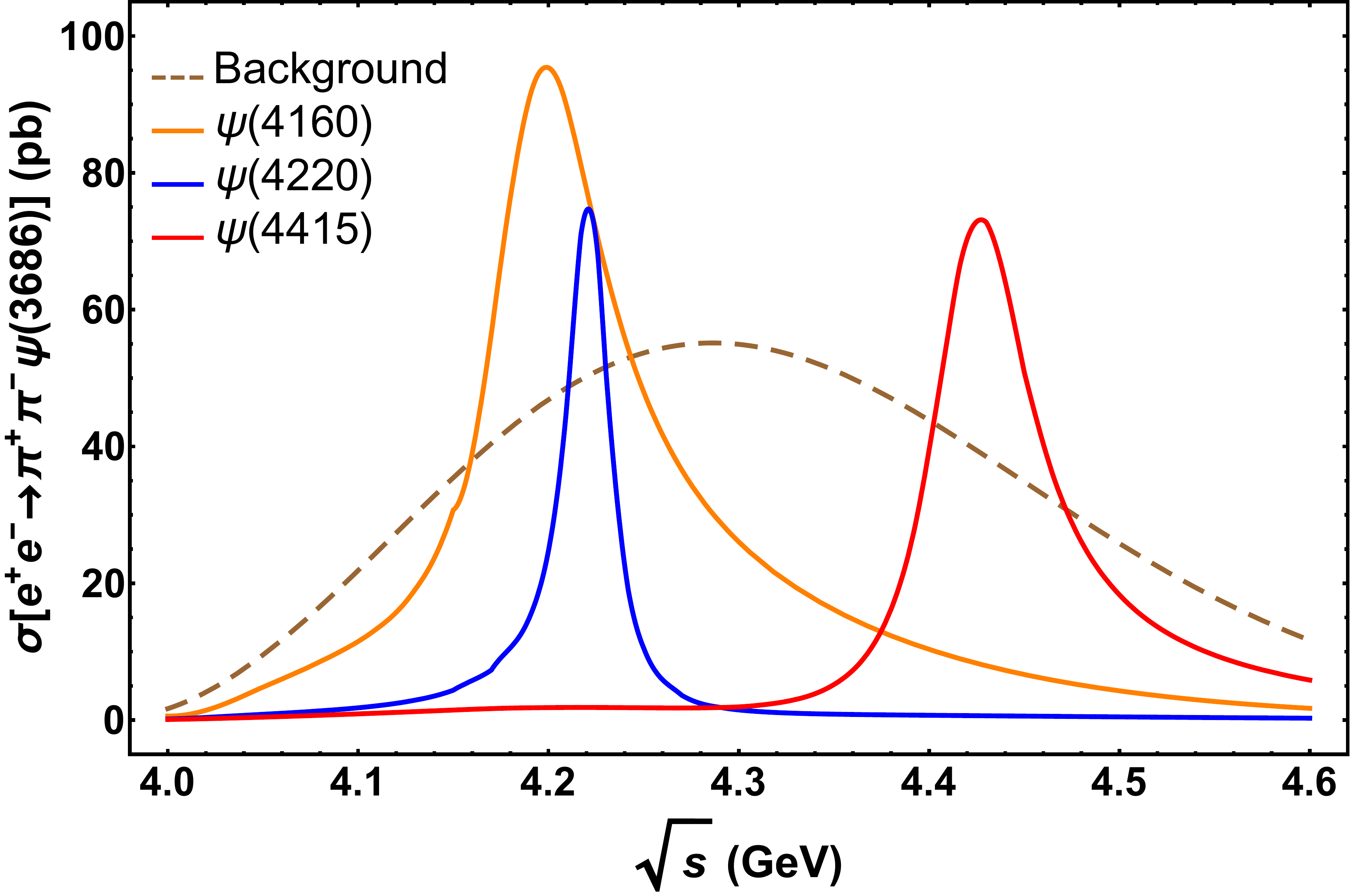}}
		\caption{(Color online). Individual contributions from the intermediate charmonia and background in the three-charmonium scenario.}
		\label{fig:3RC}
	\end{figure}
\end{center}

The fitted cross section for $e^+ e^- \to \pi^+ \pi^- \psi(3686)$ is presented in Fig. \ref{fig:cs} (black dashed curve). The structure in the vicinity of 4.2 GeV and the broad enhancement near 4.4 GeV  are well reproduced. However, one can find that the fitted curve is above the experimental data in the vicinity of 4.3 GeV.
Such a phenomenon indicates that there may exist other intermediate charmonium contributions to the $e^+e^- \to \pi^+ \pi^- \psi(3686)$ process, which inspires us to introduce a four-charmonium scenario to study the $e^+e^- \to \pi^+ \pi^- \psi(3686)$ process, which will be discussed in the next subsection.  In Fig.~\ref{fig:3RC}, we present the individual contributions from the nonresonance background and three charmonia. each of which is the coherent sum of direct contributions, the scalar productions, and ISPE processes for each resonance. One can find the contribution from $\psi(4160)$ could reach up to nearly 100 pb, while the contributions from $\psi(4220)$ and $\psi(4415)$ are about 70 pb. From Fig.~\ref{fig:3RC}, one can find that resonances are still in a Breit-Wigner form. Thus, we simulate each lineshape in Fig.~\ref{fig:3RC} with a Breit-Wigner resonance, which is in the form $12 \pi \Gamma^{e^+ e^-} \mathcal{B}(R\to \pi^+ \pi^- \psi(3686)) \Gamma_R/[(s-m_R^2)^2+m_R^2 \Gamma_R^2]$. With the dilepton decay widths listed in Table \ref{tab:psi-parameter}, one can roughly estimate the center values of the branching ratios for $\psi_R\to \pi^+ \pi^- \psi(3686)$, where $\psi_R=(\psi(4160),\ \psi(4220),\ \psi(4415))$. As shown in Table \ref{tab:branching-ratio}, one can find the center values of branching ratios are of the order of $\sim 10^{-3}$ for $\psi(4220)$ and $\psi(4415)$ and $\sim 10^{-2}$ for $\psi(4160)$.

\begin{table}[htpb]
\centering \caption{Branching ratios of  $\psi_R \to \pi^+ \pi^- \psi(3686) $ obtained from the present fit in the unit of $10^{-3}$.}\label{tab:branching-ratio}
\begin{tabular}{p{40pt}<\centering p{40pt}<\centering p{40pt}<\centering p{40pt}<\centering p{40pt}<\centering}
\toprule[1pt]
~ & $\psi(4160)$ & $\psi(4220)$ & $\psi(4415)$ & $\psi(4380)$\\
\midrule[0.6pt] %
3R Fit & 15.711   & 8.101   & 9.65  & -\\
4R Fit & 4.389  & 7.370  & 5.368  & 17.285 \\
\bottomrule[1pt]
\end{tabular}
\end{table}

\begin{center}
	\begin{figure}[htbp]
		\scalebox{0.225}{\includegraphics{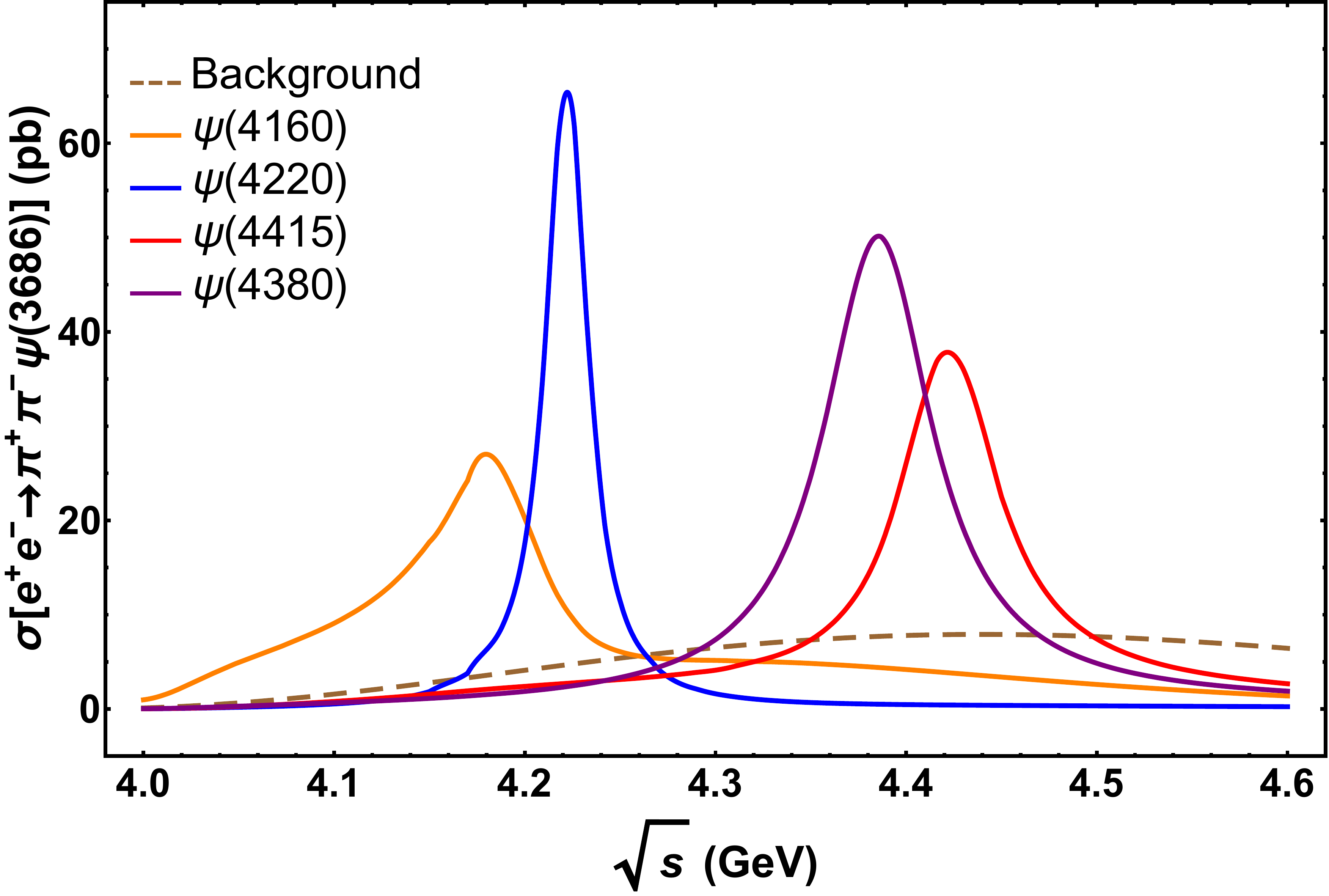}}
		\caption{(Color online). The same as Fig. \ref{fig:3RC} but in the four-charmonium scenario.}
		\label{fig:4RC}
	\end{figure}
\end{center}
\begin{table}[htpb]
\centering \caption{The same as Table \ref{tab:fit-parameter} but for the four-charmonium scenario. }\label{tab:fit-parameter-4R}
\begin{tabular}{ccccc}
\toprule[1pt]
Parameter  & $\psi(4160)$ & $\psi(4220)$ & $\psi(4415)$ & $\psi(4380)$\\
\midrule[1pt] %
$a_{R}$ (GeV$^{-1}$) & 5.5 & 4.0 & 3.3 & 3.0\\
$\phi_{\rm Dir}$ & -0.146 & 1.661 & -0.423 & -0.345\\
$F_{\rm Dir}$  & -12.243 & 3.361 & 0.504 & 1.523\\
$\kappa_{\rm Dir}$  & -0.151 & -0.215 & -1.190 & -0.463\\
$\phi_{\sigma}$ & 0.517 & -0.950 & 2.910 & -0.328\\
$f_{\sigma}$ (GeV$^2$)& 11.953 & 1.813 & 1.832 & -5.940\\
$\varphi_{\sigma}$ & -1.680 & 0.287 & -0.838 & -1.455\\
$g_{\sigma}$ & 58.717 & -6.612 & -7.864 & -25.906\\
$\phi_{ \mathrm{ISPE}}$ & -1.759 & 2.875 & 1.396 & -5.475\\
$g_{R D \bar{D} \pi}$ (GeV$^{-3}$) & -2.223 & -1.736 & -0.257 & 0.276\\
$g_{R D^\ast \bar{D} \pi}$  & 1.646 & 0.353 & -0.470 & 1.249\\
$g_{R D^\ast \bar{D}^\ast \pi}$ (GeV$^{-1}$) & 0.323 & 0.131 & 0.133 & -0.022\\
\midrule[1pt]
\multicolumn{2}{c}{$g_{\rm NoR}$=-4.847 GeV} & \multicolumn{3}{c}{$a_{\rm NoR}$=2.058 GeV$^{-2}$}\\
\midrule[1pt]
\multicolumn{5}{c}{$\chi^2/d.o.f.=2.767$}\\
\bottomrule[1pt]
\end{tabular}
\end{table}

\subsection{Four-charmonium scenario to $e^+e^- \to \psi(3686) \pi^+ \pi^-$}\label{four}

As discussed in the three-charmonium scenario in Fig.~\ref{fig:cs}, the experimental data in the vicinity of 4.3 GeV cannot be well reproduced. In Ref. \cite{Wang:2019}, the mass spectra and decay properties of higher charmonia have been investigated, and
a charmonium $\psi(4380)$ as a partner of $\psi(4220)$ was predicted, which has
the resonance parameters  $m=4384$ MeV and $\Gamma=84$ MeV. This progress stimulates us to introduce the four-charmonium scenario to study $e^+e^- \to \psi(3686) \pi^+ \pi^-$, namely, four charmonia $\psi(4160)$, $\psi(4415)$, $\psi(4220)$, and $\psi(4380)$ are considered when fitting the data. All the resonance parameters are listed in Table \ref{tab:psi-parameter}.
Indeed,
we find that the details of the cross section of $e^+e^- \to \psi(3686) \pi^+ \pi^-$ around $\sqrt{s}=4.3$ GeV
can be depicted.

The fitted parameters are presented in Table \ref{tab:fit-parameter-4R}.  With the center values of these parameters, one can get the differential cross sections for each invariant mass squared distribution and the cross section depending on $\sqrt{s}$, which are presented in Fig. \ref{fig:ims} (red curves) and Fig. \ref{fig:cs} (red curve), respectively. After including $\psi(4380)$, one finds that almost all the differential cross sections can be quantitatively fitted better than the three resonance-scenario. Especially,  the jump near at 15 GeV$^2$ in $m^2_{\pi^\pm \psi(3686) }$ invariant mass spectrum at $\sqrt{s}=4.258$ GeV can be well reproduced and $m^2_{\pi^+\pi^-}$ invariant mass spectrum at this energy point is well fitted. The fits to differential cross sections at $\sqrt{s}=4.358$ GeV are also much better than the one in the three-resonance scenario. As for the $\sqrt{s}$ dependent cross section, the four-resonance scenario can well reproduce experimental data, especially the data near at 4.3 GeV.  Quantitatively, the $\chi^2/{\rm d.o.f}$ is reduced to 2.77 in the four-charmonium scenario. Similar to the case of the three-charmonium scenario, we present the individual contributions of the non-resonance background and resonances in Fig. \ref{fig:4RC}. One can find that the resonance contributions are smaller than the corresponding one in the three-charmonium scenario. The fitted branching ratios of $\psi_R\to \pi^+ \pi^- \psi(3686)$ are presented in Table  \ref{tab:branching-ratio}, where the central values of these branching ratios for $\psi(4160)$, $\psi(4220)$, and $\psi(4415)$ are of the order of $\sim10^{-3}$, while the one for $\psi(4380)$ is of the order of $\sim10^{-2}$. 

\section{Summary}
\label{Sec:Sum}

In the past years, the BESIII and Belle Collaborations have made a great progress on finding charged charmonium-ike $XYZ$ states.
A series of charged charmonium-like structures, $Z_c(3900)$ \cite{Ablikim:2013mio},  $Z_c(3885)$ \cite{Ablikim:2013xfr}, $Z_c(4020)$ \cite{Ablikim:2013wzq}, and $Z_c(4025)$ \cite{Ablikim:2013emm}, were reported one by one.
These experimental results have stimulated a great interest of both theorists and experimentalists since the charged property of these charmonium-like states indicates these states should contain at least four constituent quarks, thus they could be good candidates of tetraquark states. The estimates in frameworks of the QCD sum rule \cite{Qiao:2013dda, Wang:2013vex,Wang:2013exa, Wang:2013llv, Agaev:2017tzv,  Zhao:2014qva}  and potential model \cite{Zhu:2016arf, Deng:2014gqa, Anwar:2018sol, Patel:2014vua, Liu:2017mrh} indicate that both $Z_c(3900)/Z_c(3885)$ and $Z_c(4020)/Z_c(4025)$ could be explained as tetraquark states with different configurations.  Furthermore, the observed masses of $Z_c(3900)/Z_c(3885)$ and $Z_c(4020)/Z_c(4025)$ are very close to the thresholds of $D^\ast \bar{D}$ and $D^\ast \bar{D}^\ast$, thus, these charmonium-like states have been extensively investigated in deutron-like hadronic molecular scenarios. Based on the molecular assumption, mass spectrum, decay properties, and productions of these charmonium-like states have been investigated in different models, such as QCD sum rule \cite{Cui:2013yva, Zhang:2013aoa, Cui:2013vfa, Wang:2013daa}, potential model \cite{Liu:2008tn, Liu:2008fh, Sun:2012zzd, He:2013nwa},  and some other phenomenological Lagrangian approaches \cite{Gutsche:2014zda, Chen:2015igx, Li:2013xia, Li:2014pfa, Xiao:2018kfx, Chen:2016byt}.

However, before definitely identifying these charged charmomium-like structures as exotic tetraquark states,
we need to exhaust all the possibilities of explaining them in the conventional theoretical framework. Here, we must mention our previous prediction of charged charmonium-like structures in our theoretical work \cite{Chen:2011xk}, where these predicted charged charmonium-like structures by the ISPE mechanism may exist in the $J/\psi\pi^\pm$, $h_c\pi^\pm$, and $\psi(3686)\pi^\pm$ invariant mass spectra of higher charmonia, and in the charmonium-like state $Y(4260)$ decays into $J/\psi\pi^+\pi^-$, $h_c\pi^+\pi^-$ and $\psi(3686)\pi^+\pi^-$ \cite{Chen:2011xk}. After two years of our paper,
the BESIII and Belle collaborations discovered a charged charmonium-like $Z_c(3900)$ in the process $e^+ e^- \to J/\psi \pi^+ \pi^-$ around $\sqrt{s}=4.26\ \mathrm{GeV}$ \cite{Ablikim:2013mio,Liu:2013dau}, and  CLEO-c confirmed it in the same process but at $\sqrt{s}=4.17$ GeV \cite{Xiao:2013iha}. In 2013, BESIII observed another charged charmoniumlike structure named as $Z_c(4020)$ in the $h_c \pi^+$ invariant mass spectrum of the $e^+ e^- \to \pi^+ \pi^- h_c$ process \cite{Ablikim:2013wzq}. These experimental observations make us believe that the ISPE mechanism indeed plays an important role in producing these novel phenomena. What is more important is that these measured experimental data make us possible to study them with higher precision. In Ref. \cite{Chen:2013coa}, we fitted the experimental result of the charged $Z_c(3900)$ observed in the $J/\psi\pi^\pm$ invariant mass spectrum \cite{Ablikim:2013mio,Liu:2013dau} and indicated that the charged $Z_c(3900)$ structure can be well established based on  the ISPE mechanism. This study further enforces theorist's confidence to explain these experimental results in the conventional theoretical framework. Similar studies were proposed in Refs. \cite{Swanson:2014tra,Szczepaniak:2015eza,Swanson:2015bsa,Bugg:2011jr,Liu:2015taa}. Later, the Lattice QCD calculation \cite{Ikeda:2017mee, Ikeda:2016zwx} also supports such a scenario.

In 2017, the BESIII Collaboration again observed charged charmonium-like structures in the $\psi(3686)\pi^\pm$ invariant mass spectrum of the $e^+e^-\to \psi(3686)\pi^+\pi^-$ process at different energy points $\sqrt{s}=4.226,\,4.258,\,4.358,\,4.387,\,4.416,\,4.600$ GeV \cite{Ablikim:2017oaf}.
This new observation inspires our interest in further testing the ISPE mechanism combined with the present precise experimental data.
A crucial task of this work is to examine whether the cross section of the $e^+e^-\to \psi(3686)\pi^+\pi^-$ process, and the corresponding $\psi(3686)\pi^\pm$ and $\pi^+\pi^-$ invariant mass spectra at different $\sqrt{s}$ values can be reproduced by the ISPE mechanism. As illustrated by our calculation, we find that the reported charged charmonium-like structures in the $\psi(3686)\pi^\pm$ invariant mass spectrum of the $e^+e^-\to \psi(3686)\pi^+\pi^-$ process at different energy points can be depicted in a unified framework, which sheds light on the properties of the observed charged charmonium-like structures. We need to emphasize that our study also further supports the existence of a new higher charmonium $\psi(4380)$ which was predicted in Ref. \cite{Wang:2019}, since the fitting result under the four-charmonium scenario is better than that under the three-charmonium scenario. We strongly suggest the BESIII and BelleII collaborations to search for this predicted missing charmonium $\psi(4380)$, especially by analyzing  the $e^+e^-\to \psi(3686)\pi^+\pi^-$ process in near future.

In the following years, studies on charmonium-like $XYZ$ states will still be an interesting research topic at the typical facilities like BESIII, LHCb, and BelleII. Since the observed charmonium-like $XYZ$ states can be a good candidate of an exotic tetraquark matter, it is a main task for both experimentalists and theorists how to identify them as a genuine exotic multiquark state. To achieve this aim, we need to check whether the charmonium-like $XYZ$ states can be assigned into a conventional charmonium family or can be settled down in a conventional theoretical framework. The present work is the effort along this line. Besides theoretical investigation, the Lattice QCD study about charmonium-like $XYZ$ will provide  valuable information to shed light on the underlying properties of the $XYZ$ states.
As emphasized in a recent review article \cite{Liu:2019zoy}, a joint effort from phenomenological method, experimental  analysis, and Lattice QCD calculation should be paid more attentions on, which must promote our knowledge of how to form charmonium-like $XYZ$ states.

\section*{Acknowledgements}

This work is partly supported by the China National Funds for Distinguished Young Scientists under Grant No. 11825503, the National Natural Science Foundation of China under Grant No. 11775050, National Program for Support of Top-notch Young Professionals, and the Fundamental Re- search Funds for the Central Universities.

\appendix
\section{Amplitudes of the ISPE mechanism}
\label{Sec:App-Amp}

\begin{center}
\begin{figure}[htbp]
  \begin{tabular}{cc}
  \scalebox{0.5}{\includegraphics{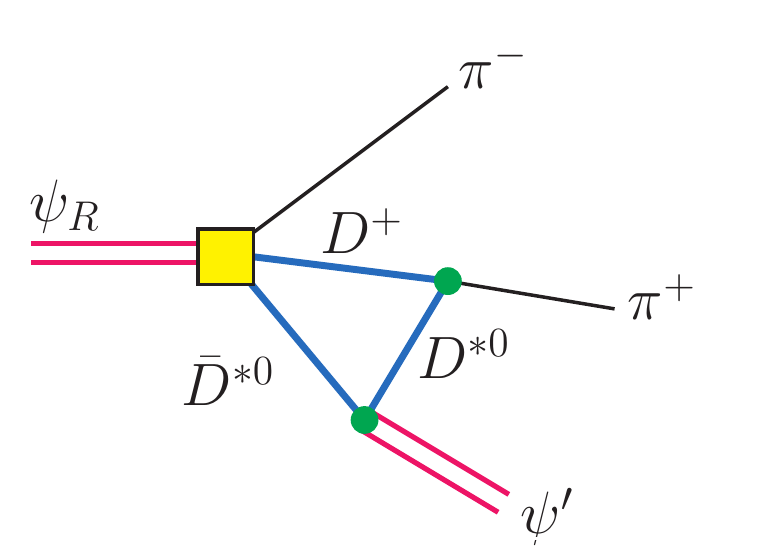}} &
  \scalebox{0.5}{\includegraphics{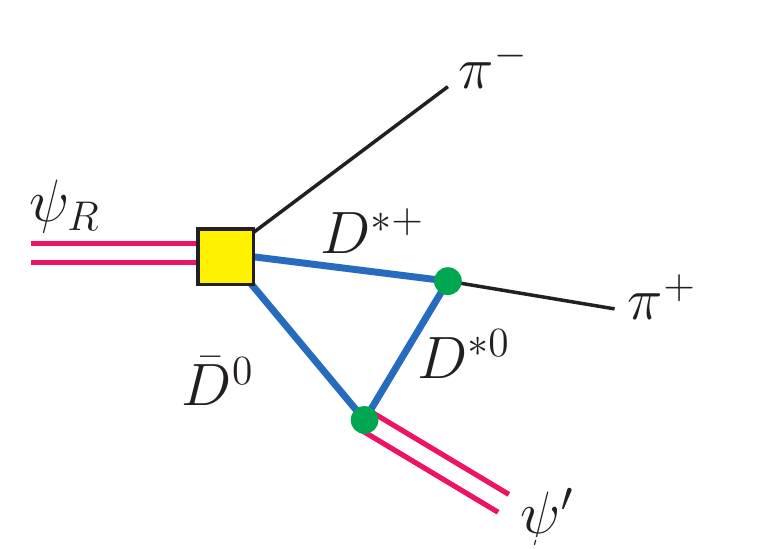}} \\
  \large$(a)$ & \large$(b)$\\
  \scalebox{0.5}{\includegraphics{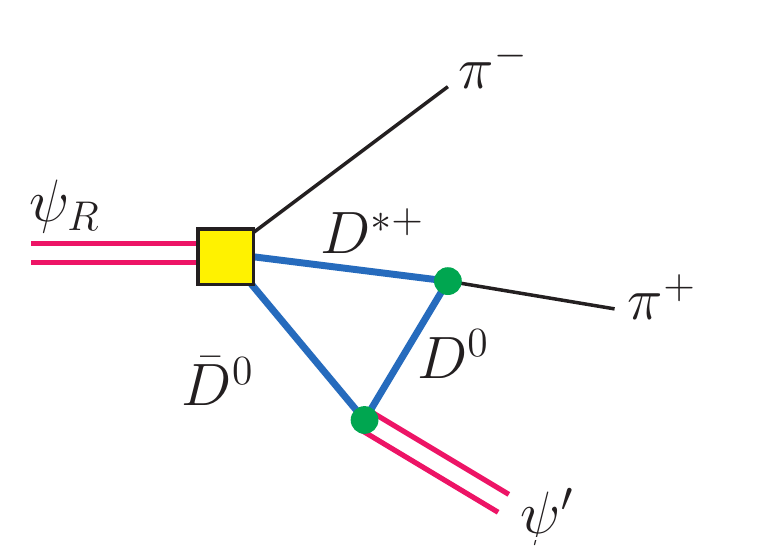}} &
  \scalebox{0.5}{\includegraphics{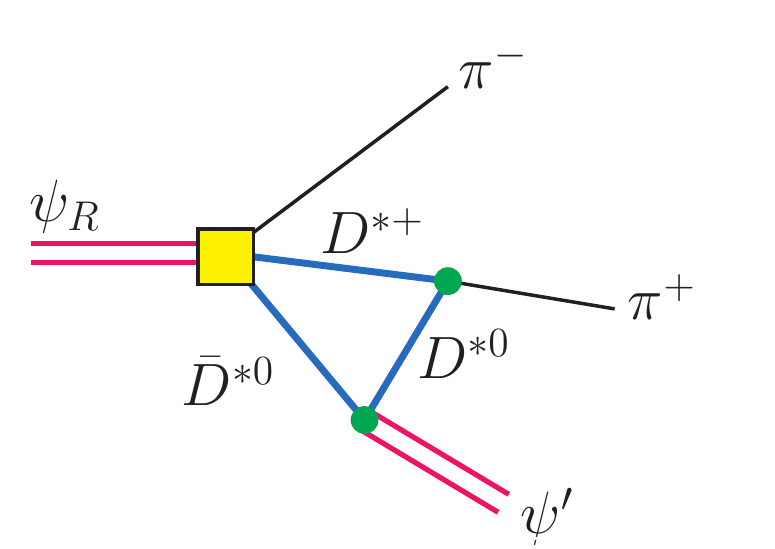}}\\
  \large$(c)$ & \large$(d)$\\
  \scalebox{0.5}{\includegraphics{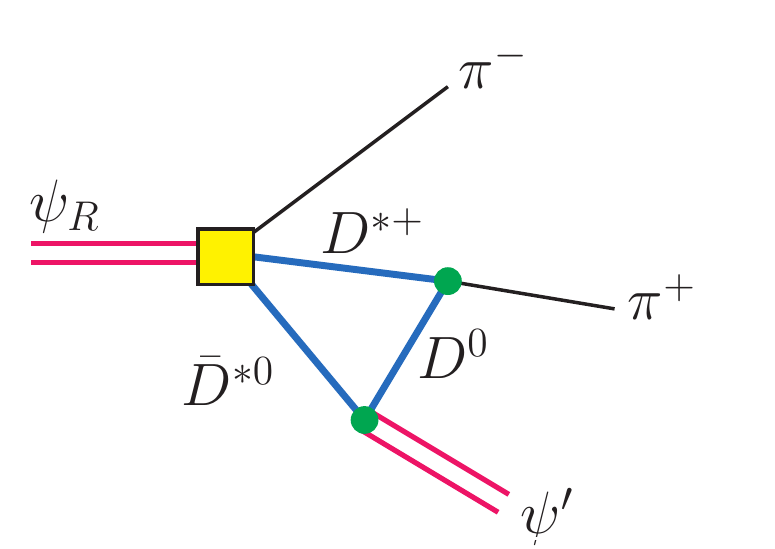}}&
  \scalebox{0.5}{\includegraphics{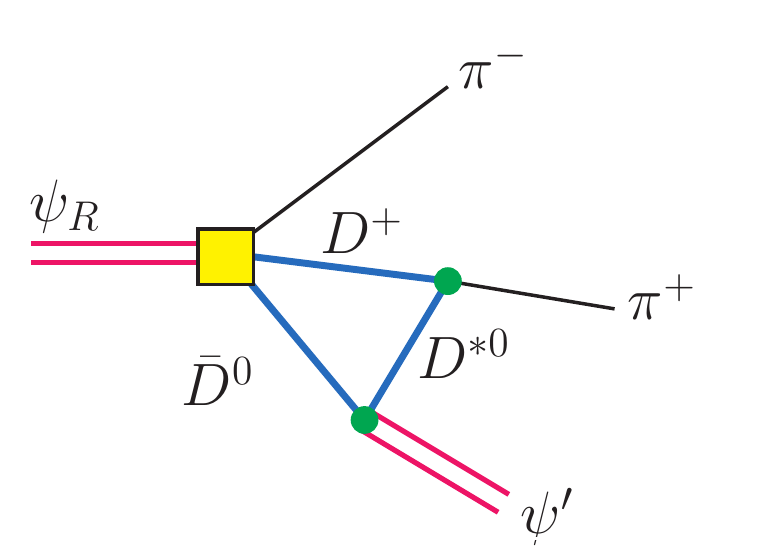}}\\
  \large$(e)$ & \large$(f)$
  \end{tabular}
  \caption{(Color online). Typical diagrams of the ISPE mechanism with a $\pi^-$ emission at the hadronic level. Diagrams $(a)-(c)$, $(d)-(e)$ and $(f)$ are the contributions from $D^\ast \bar{D} +h.c.$, $D^\ast \bar{D}^\ast$ and $D\bar{D}$ intermediate processes, respectively.  }
  \label{fig:ISChE-BBS-13}
\end{figure}
\end{center}

In this Appendix, we will present some details of the ISPE mechanism. Under the ISPE mechanism, $\psi_R$ transits into $D^{(\ast)}$ and $\bar{D}^{(\ast)}$ pair associated with a single pion emission. Since the emitted pion has continuous energy distribution, $D^{(\ast)}$ and $\bar{D}^{(\ast)}$ mesons with the low momentum can easily interact with each other and further transit into $\psi(3686)\pi^\pm$ by exchanging a $D^{(\ast)}$ meson. In Fig. \ref{fig:ISChE-BBS-13}, we present some typical diagrams of the ISPE mechanism with a $\pi^-$ emission at the hadronic level. For convenience, in these diagrams, we denote $\psi(3686)$ as $\psi^\prime$. Although there are only 6 diagrams in Fig. \ref{fig:ISChE-BBS-13}, to get the whole contribution of the ISPE mechanism, we must consider the diagrams caused by isospin symmetry and initial $\pi^+$ emission. For isospin symmetry, the diagrams can be obtained by the replacement $\{ D^{(\ast)+}, D^{(\ast) 0}, \bar{D}^{(\ast) 0}\} \to \{\bar{D}^{(\ast)0}, D^{(\ast) -}, D^{(\ast) +}\}$, which are the same contributions as for a $\pi^-$ emission. For contributions of an initial $\pi^+$ emission, the typical diagrams of the ISPE mechanism can be obtained by applying charge conjugation on each Fig. \ref{fig:ISChE-BBS-13} (i) (i=a$\sim$f). Thus, the total decay amplitude can be expressed as
\begin{eqnarray}
\mathcal{A}_{\mu\zeta} &=& 2 \sum\limits_{i=a}^f \Big(\mathcal{A}_{\mu\zeta}^{(i)}+\mathcal{A}_{\mu\zeta[{p_2 \rightleftharpoons p_3}]}^{(i)} \Big),\label{eqs:ISPE-total-A}
\end{eqnarray}
where the factor 2 reflects the isospin symmetry mentioned above, and $\mathcal{A}_{\mu\zeta[{p_2 \rightleftharpoons p_3}]}^{(i)}$ means that the contributions of an initial $\pi^+$ emission can be obtained by applying $p_2 \rightleftharpoons p_3$ transformation on $\mathcal{A}_{\mu\zeta}^{(i)}$.

To write the decay amplitudes $\mathcal{A}_{\mu\zeta}^{(i)}$ in Eq. (\ref{eqs:ISPE-total-A}), the effective Lagrangian approach is adopted. For the four particles interaction, the effective Lagrangian is \cite{Chen:2011pv,Oh:2000qr}
\begin{eqnarray}
&&\mathcal{L}_{\psi D^{(*)} D^{(*)} \pi}\nonumber\\&&=-ig_{R DD \pi}
\varepsilon^{\mu \nu \alpha \beta} \psi_{\mu} \partial_{\nu} D
\partial_{\alpha} \pi \partial_{\beta} \bar{D} + g_{R D^\ast D \pi} \psi^{\mu} (D \pi
\bar{D}^\ast_{\mu} + D^\ast_{\mu} \pi \bar{D}) \nonumber\\
&&\quad-ig_{R D^\ast D^\ast \pi} \varepsilon^{\mu \nu \alpha
\beta} \psi_{\mu} D^\ast_{\nu} \partial_{\alpha} \pi
\bar{D}^\ast_\beta -ih_{R D^\ast D^\ast \pi} \varepsilon^{\mu \nu \alpha
\beta} \partial_{\mu} \psi_{\nu} D^\ast_{\alpha} \pi
\bar{D}^\ast_{\beta}.\label{eqs:psiDDpi}\nonumber\\
\end{eqnarray}
We want to note here that, to reduce the number of free parameters in our fit, we assume $h_{R D^\ast D^\ast \pi}=g_{R D^\ast D^\ast \pi}$ in this work.

The effective Lagrangian that describes the interaction among a pair of charmed mesons and a pion is \cite{Chen:2011pv,Casalbuoni:1996pg,Colangelo:2003sa}
\begin{eqnarray}
&&\mathcal{L}_{D^\ast D^{(\ast)} \pi} \nonumber\\&&= ig_{D^\ast D \pi}
(D^\ast_{\mu} \partial^\mu \pi \bar{D}-D \partial^\mu \pi
\bar{D}^\ast_{\mu})-g_{D^\ast D^\ast \pi} \varepsilon^{\mu \nu \alpha \beta}
\partial_{\mu} D^\ast_{\nu} \pi \partial_{\alpha}
\bar{D}^\ast_{\beta}.\label{eqs:DDpi}\nonumber\\
\end{eqnarray}

Finally, the interaction between $D^{(\ast)}$, $\bar{D}^{(\ast)}$, and $\psi$ can be described as \cite{Chen:2011pv,Casalbuoni:1996pg,Colangelo:2003sa}
\begin{eqnarray}
&&\mathcal{L}_{\psi D^{(*)} D^{(*)}}\nonumber\\&&=ig_{\psi DD}
\psi_{\mu} (\partial^\mu D \bar{D}- D \partial^\mu
\bar{D})-g_{\psi D^\ast D} \varepsilon^{\mu \nu \alpha \beta}
\partial_{\mu} \psi_{\nu} (\partial_{\alpha} D^\ast_{\beta} \bar{D}
\nonumber\\&&\quad+ D \partial_{\alpha} \bar{D}^\ast_{\beta})-ig_{\psi D^\ast D^\ast} \big\{ \psi^\mu (\partial_{\mu}
D^{\ast \nu} \bar{D}^\ast_{\nu} -D^{\ast \nu} \partial_{\mu}
\bar{D}^\ast_{\nu}) \nonumber\\
&&\quad+ (\partial_{\mu} \psi_{\nu} D^{\ast \nu} -\psi_{\nu}
\partial_{\mu} D^{\ast \nu}) \bar{D}^{\ast \mu} + D^{\ast \mu}(\psi^\nu \partial_{\mu} \bar{D}^\ast_{\nu} -
\partial_{\mu} \psi^\nu \bar{D}^\ast_{\nu})\big\}.\label{eqs:DDpsi}\nonumber\\
\end{eqnarray}

For the coupling constants $g_{D^{(\ast)}D^{(\ast)}\pi}$ in Eq. (\ref{eqs:DDpi}) and $g_{\psi D^{(\ast)}D^{(\ast)}}$ in Eq. (\ref{eqs:DDpsi}), heavy quark symmetry gives the relations between them separately, which can be written as
\begin{eqnarray}
&&\frac{2g_\pi}{f_\pi} = g_{D^\ast D^\ast \pi} = \frac{g_{D^\ast D \pi}}{\sqrt{m_D m_{D^*}}},\\
&&g_{\psi DD}=g_{\psi D^\ast D^\ast} \frac{m_D}{m_{D^\ast}}=g_{\psi D^\ast D} \sqrt{m_D m_{D^\ast}}=\frac{m_{\psi}}{f_{\psi}}.\label{eqs:cc}
\end{eqnarray}
In Eq. (\ref{eqs:cc}), ${g_\pi = 0.59}$, ${f_\pi = 0.132}$ GeV, and the decay constant of $\psi(3686)$ is $f_{\psi}=0.416$.

With above effective Lagrangians, the expressions for decay amplitudes $\mathcal{A}_{\mu\zeta}^{(i)}$ can be written as
\begin{eqnarray}
\mathcal{A}_{\mu\zeta}^{(a)} &=& (i)^3 \int \frac{d^4q}{(2\pi)^4} [g_{RD\bar{D}^*\pi}] [-i g_{DD^*\pi} (ip_3^\lambda)]\nonumber\\
&&\times [g_{\bar{D}^* D^* \psi^\prime} \left(g_{\zeta\xi} q_\kappa - g_{\zeta\kappa} k_{2\xi} - g_{\kappa\xi} (q_\zeta-k_{2\zeta})\right)]\nonumber\\
&&\times\frac{1}{k_1^2-m_{D}^2} \frac{-g_{\mu}^{\xi} + k_{2\mu} k_{2}^{\xi}/m_{D^*}^2}{k_2^2-m_{D^*}^2} \frac{-g_{\lambda}^{\kappa} + q_\lambda q^\kappa/m_{D^*}^2}{q^2-m_{D^*}^2} \mathcal{F}^2(q^2),\nonumber\\
\mathcal{A}_{\mu\zeta}^{(b)} &=& (i)^3 \int \frac{d^4q}{(2\pi)^4} [g_{RD^*\bar{D}\pi}] [-g_{D^*D^*\pi} \varepsilon^{\lambda\rho\delta\sigma} (-i k_{1\lambda}) (iq_\delta)]\nonumber\\
&&\times [g_{\bar{D} D^* \psi^\prime} \varepsilon_{\zeta\eta\kappa\xi} p_{3}^{\eta} (q^\xi-k_{2}^{\xi})]\frac{-g_{\mu\rho} + k_{1\mu} k_{1\rho}/m_{D^*}^2}{k_1^2-m_{D^*}^2}\nonumber\\
&&\times  \frac{1}{k_2^2-m_{D}^2} \frac{-g_{\sigma}^{\kappa} + q_\sigma q^\kappa/m_{D^*}^2}{q^2-m_{D^*}^2} \mathcal{F}^2(q^2),\nonumber\\
\mathcal{A}_{\mu\zeta}^{(c)} &=& (i)^3 \int \frac{d^4q}{(2\pi)^4} [g_{RD^*\bar{D}\pi}] [i g_{D^*D\pi} (ip_3^\lambda)][g_{\bar{D} D \psi^\prime} {\psi^\prime}(q_\zeta-k_{2\zeta})]\nonumber\\
&&\times \frac{-g_{\mu\lambda} + k_{1\mu} k_{1\lambda}/m_{D^*}^2}{k_1^2-m_{D^*}^2} \frac{1}{k_2^2-m_{D}^2} \frac{1}{q^2-m_{D}^2} \mathcal{F}^2(q^2),\nonumber
\end{eqnarray}
\begin{eqnarray}
\mathcal{A}_{\mu\zeta}^{(d)} &=& (i)^3 \int \frac{d^4q}{(2\pi)^4} [i g_{RD^*\bar{D}^*\pi}\varepsilon_{\mu\nu\alpha\beta} (ip_{1}^{\alpha}-ip_{2}^{\alpha})]\nonumber\\
&&\times [-g_{D^*D^*\pi} \varepsilon^{\lambda\rho\delta\sigma} (-i k_{1\lambda}) (iq_\delta)]\nonumber\\
&&\times [g_{\bar{D}^* D^* \psi^\prime} \left(g_{\zeta\xi} q_\kappa - g_{\zeta\kappa} k_{2\xi} - g_{\kappa\xi} (q_\zeta-k_{2\zeta})\right)]\nonumber\\
&&\times \frac{-g^{\nu}_{\rho} + k_{1}^{\nu}k_{1\rho}/m_{D^*}^2}{k_1^2-m_{D^*}^2} \frac{-g^{\beta\xi} + k_{2}^{\beta}k_{2}^{\xi}/m_{D^*}^2}{k_2^2-m_{D^*}^2}\nonumber\\
&&\times \frac{-g_{\sigma}^{\kappa} + q_\sigma q^\kappa/m_{D^*}^2}{q^2-m_{D^*}^2} \mathcal{F}^2(q^2),\nonumber\\
\mathcal{A}_{\mu\zeta}^{(e)} &=& (i)^3 \int \frac{d^4q}{(2\pi)^4} [i g_{RD^*\bar{D}^*\pi}\varepsilon_{\mu\nu\alpha\beta} (ip_{1}^{\alpha}-ip_{2}^{\alpha})]\nonumber\\
&&\times [i g_{D^*D\pi} (ip_3^\lambda)] [g_{\bar{D}^* D \psi^\prime} \varepsilon_{\zeta\eta\kappa\xi} p_{3}^{\eta} (k_{2}^{\xi}-q^\xi)]\nonumber\\
&&\times \frac{-g^{\nu}_{\lambda} + k_{1}^{\nu}k_{1\lambda}/m_{D^*}^2}{k_1^2-m_{D^*}^2} \frac{-g^{\beta\kappa} + k_{2}^{\beta}k_{2}^\kappa/m_{D^*}^2}{k_2^2-m_{D^*}^2}\nonumber\\
&&\times \frac{1}{q^2-m_{D}^2} \mathcal{F}^2(q^2),\nonumber
\end{eqnarray}
\begin{eqnarray}
\mathcal{A}_{\mu\zeta}^{(f)} &=& (i)^3 \int \frac{d^4q}{(2\pi)^4} [-i g_{RD\bar{D}\pi} \varepsilon_{\mu\nu\alpha\beta} (ik_{1}^{\nu}) (ip_{2}^{\alpha}) (ik_{2}^{\beta})]\nonumber\\
&&\times [-i g_{DD^*\pi} (ip_3^\lambda)] [g_{\bar{D} D^* \psi^\prime} \varepsilon_{\zeta\eta\kappa\xi} p_{3}^{\eta} (q^\xi-k_{2}^{\xi})]\nonumber\\
&&\times \frac{1}{k_1^2-m_{D}^2} \frac{1}{k_2^2-m_{D}^2} \frac{-g_\lambda^\kappa + q_\lambda q^\kappa/m_{D^*}^2}{q^2-m_{D^*}^2} \mathcal{F}^2(q^2).\nonumber
\end{eqnarray}

In $\mathcal{A}_{\mu\zeta}^{(i)}$, a monopole form factor $\mathcal{F}(q^2) = \frac{q^2-\Lambda^2}{m_E^2-\Lambda^2} $ is introduced to reflect the structure effect at each vertex and the off-shell effect of the exchanged $D^{(\ast)}$ meson. In this form factor, $m_E$ is the mass of the exchanged $D^{(\ast)}$ meson, $\Lambda$ is parameterized as $\Lambda=m_E+\alpha_\Lambda \Lambda_{QCD}$ with $\Lambda_{QCD}=220$ MeV. Since Ref. \cite{Chen:2011xk} has shown that the ISPE mechanism is weakly dependent on $\alpha_\Lambda$, thus in this work we also set $\alpha_\Lambda=1$ to present our results.


\end{document}